\documentclass[showpacs,preprintnumbers,amsmath,amssymb]{revtex4}%
\begin{document}%

\typeout{Filename: reftest4-1.tex for revtex 4.1i 2009/10/19 (AO)}

{\Large
 {Intertwining of exactly solvable generalized Schr\"odinger
equations} }
\\
\author
 {\large
{A.A. Suzko$^{a,b}$, E.P.Velicheva${}^a$} }
\address{$^{a}$Joint Institute for Nuclear Research, 141980 Dubna,
Russia}
\address{ ${}^b$ JIPENR, National Academy of Sciences of
Belarus, Minsk}
\begin{abstract}
The Darboux transformation operator technique in differential and
integral forms is applied to the generalized Schr\"odinger equation
with a position-dependent effective mass and with linearly
energy-dependent potentials. Intertwining operators are obtained in
an explicit form and used for constructing  generalized Darboux
transformations of an arbitrary order. A relation between
supersymmetry and the generalized Darboux transformation is
considered. The method is applied to generation of isospectral
potentials with additional or removal bound states or construction
of new partner potentials without changing the spectrum, i.e. fully
isospectral potentials. The method is illustrated by some examples.
\end{abstract}

\maketitle

\section{Introduction}
The problem of exact solvability of the  Schr\"odinger equation has
been extensively considered since the beginning of quantum mechanics
\cite{Schr}. A factorization technique was introduced by
Schr\"odinger and was used to solve the harmonic oscillator, the
hydrogen  atom, the Kepler motion. It should be noted that
factorization is closely connected with the Darboux transformation
\cite{Darboux} and the supersymmetry approach introduced by Witten
in quantum mechanics \cite{Witten}. Many interesting exactly
solvable models have been constructed for Schr\"odinger equation
using the Darboux and Bargmann transformation techniques
\cite{Bagrov}-\cite{Suz93}.
 Many advances have been made in the field of classifying quantum mechanical
potentials according to their symmetry properties and in the area of
different applications of exactly solvable models in quantum
mechanics, in particular, in atomic and nuclear physics, in
statistical mechanics \cite{Junker}-\cite{Suz93}. Potentials of
Bargmann- type and corresponding exact solutions can be obtained
from the integral equations of the inverse scattering problem in the
Gelfand-Levitan \cite{Gelfand} and Marchenko approaches
\cite{Marchenko} with the degenerate kernels of the transformation
operator. The differential Darboux transformations (or the method of
itertwining) are closely related to supersymmetry and have a variety
of applications in different fields of physics. The technique of
iterated integral operators for Schr\"odinger equation is based on
the integral Gelfand-Levitan and Marchenko equations  (see e.g.
\cite{Levitan}). The differential and integral approaches  of
obtaining exactly solvable equations are very close to each other,
but sometimes the differential method is more convenient  and
sometimes  the integral method is preferable. In particular, the
integral form of transformations is used for construction of
phase-equivalent potentials. Suggested in \cite{Scripta84,Suz85}
generalized Darboux and Bargmann transformations permit construction
of exactly
 solvable potentials for variable values of energy ${\cal
E}$ and angular momentum $l$. The transformations in differential
and integral forms were considered and it was shown how to construct
the phase-equivalent potentials.
 Later on, this approach  was
generalized on the Schr\"odinger equation with weighted energy
\cite{Suz93}
 and recently \cite{Acta} the intertwining technique has been applied
 to construct exactly solvable potentials
for this  equation. The Schr\"odinger equation with energy-dependent
potentials is widely used in nuclear physics, helium clusters and
metal clusters \cite{Feshbach}-\cite{Arias}.

In the last few years the research efforts on the topic of algebraic
methods have been considerably intensified
\cite{Morrow}-\cite{gener} due to the rapid development of
nanoelectronics, the basic elements of which are low-dimensional
structures such as quantum wells, wires, dots and superlattices
\cite{Goser,physE}. For investigation of nonuniform semiconductors,
in which the carrier effective mass depends on position,  the
generalized Schr\"odinger equation with position-dependent effective
mass is used \cite{Bastard}. One of the most important problems of
quantum engineering is the construction of multi-quantum well
structures possessing desirable spectral properties. The  technique
of intertwining relations or Darboux transformations allows one to
model quantum well potentials with a given spectrum. Although
Darboux transformations are applicable to the position-dependent
mass Schr\"odinger equation \cite{Plastino,Milanov,suz-shul,Roy2},
the Schr\"odinger equation with weighted energy \cite{Acta}
 and the generalized Schr\"odinger equation with position-dependent mass and weighted
 energy \cite{gener,ss-JPA},
 however, their applications are very complicated for
realization and there remain many problems to study. Nowadays there
is a great interest to generate Hamiltonians with a prescribed
energy spectrum.  However, these problems are worth investigated for
all cases of the generalized Schr\"odinger equations except for the
standard Schr\"odinger equation ( see e.g.
\cite{Chadan}-\cite{matveev}).

In this  paper we  will focus on Darboux transformations of an
arbitrary order applied to  the generalized Schr\"odinger equation
with a position-dependent mass and energy-dependent potentials.
 These generalized Darboux transformations include all
known cases: Darboux transformations for the Schr\"odinger equation
with the position-dependent mass, the case with linearly
energy-dependent potentials
 as
well as Darboux transformations for the conventional Schr\"odinger
equation.
 It is known that Darboux transformations and
supersymmetry are recovered for the standard Schr\"odinger equation.
Here, we consider interrelations between supersymmetry and Darboux
transformations for the generalized Schr\"odinger equation,
establish a correspondence between the spaces of solutions to the
initial and transformed equations. We show how the procedure can be
used for generating families of Hamilltonians with a predetermined
spectrum, removing or adding new bound states. Darboux
transformations for the generalized Schr\"odinger equation are
considered in the differential and integral forms. It should be
noted,  the integral transformations can be important for generation
of  new  potentials completely isospectral to a given initial one
without changing the spectrum and for construction of Hamilltonians
differing by one bound state. A correspondence between the first-
and second-order differential Darboux transformations and the
integral ones
 is established.
 To explain our findings, we consider a few different applications of our
 approach
 and show how it is possible to generate completely isospectral potentials or to construct potentials
with addition or moving of   bound states from the spectrum of the
initial Hamiltonian.
 We analyze the influence of the distances  between the energy levels on
the shape of potentials, which can be very complex. We construct
 asymmetric double well and even triple well potentials for the
effective mass Schr\"odinger equation.

The paper is organized as follows.  Section 2 is devoted to the
generalized Darboux transformation of the first-order and
corresponding supersymmetry formalism. In Section 3 we construct
chains of generalized Darboux transformations by iteration of
first-order Darboux transformations. Furthermore, we derive
relations for potentials and solutions, obtained within the higher
order Darboux transformations, in terms of potentials and solutions
of the initial equations with no use of ones for the intermediate
equations.
 In Section 4 we derive  transformations in an integral form and
establish relationship with differential Darboux transformations,
afterwards we consider totally  isospectral potentials. In Section 5
we illustrate our generalized transformations by concrete examples.

\section{The intertwining relations and supersymmetry}
As it is well known, Darboux transformations (or supersymmetry) in
nonrelativistic quantum mechanics allows one to produce Hamiltonians
whose spectra can differ in  one bound state. Recently \cite{gener},
we have considered the first-order Darboux transformation for the
Schr\"odinger equation with a position-dependent mass and weighted
energy
\begin{equation} \label{1}
  -  \left[\frac{d}{dx}\left(\frac{1}{m(x)}\right)\frac{d}{dx}\right]\phi(x)+
v(x)\phi(x)=q(x){\cal E}\phi(x)~.\end{equation} Here $m(x)$ stands
for the particle's effective mass, $q(x)$ and $v(x)$ denote the
potentials, $\phi(x)$ is the wave function and ${\cal E}$ denotes
the real-valued energy and we  use atomic units. In this section we
will briefly describe the  method of intertwining operators of the
first-order,  then we will analyze the supersymmetry and the $n$th
order transformations.

\subsection{The first-order Darboux transformations}
 Consider two generalized Schr\"odinger
equations
\begin{eqnarray} \label{H}
&&{\cal H} \phi={\cal E}\phi~,
~~~{\cal H}=-\frac{1}{q(x)}
\left[\frac{d}{dx}\left(\frac{1}{m(x)}\right)\frac{d}{dx}\right]+
\frac{v(x) }{q(x)}~,
\\
&&\widetilde{\cal H} \widetilde\phi={\cal E}\widetilde\phi~,
\label{H-1} ~~~\widetilde{\cal H}=-\frac{1}{q(x)}
\left[\frac{d}{dx}\left(\frac{1}{m(x)}\right)\frac{d}{dx}\right]+
\frac{\widetilde{v}(x)}{q(x)}~,
\end{eqnarray}
where Hamiltonians ${\cal H}$ and $\widetilde{\cal H}$ differ only
in potentials $v$ and $\widetilde{v}$. In the conventional
intertwining technique for the Schr\"odinger equation two
one-dimensional Hamiltonians ${\cal H}=-\partial_{xx}+v$ and
$\widetilde{\cal H}=-\partial_{xx}+\widetilde{v}$ and corresponding
solutions $\phi$ and $\widetilde{\phi}$ are related by means of
differential operator ${\cal L}$
\begin{eqnarray}\label{inter1} && {\cal L}{\cal
H}=\widetilde{\cal H}{\cal L}~,
\\
&& \label{inter2}\widetilde{\phi}~=~{\cal L}~\phi~.~~~~~~
\end{eqnarray}
As it  is known, the method of intertwining differential operators
provides an universal approach to generating new exactly solvable
equations \cite{shabat,pavlov}.
 We
 expand the intertwining relations on our generalized equation
 in order to  determine the intertwining  operator ${\cal
L}$, a new potential and solutions of transformed equation
(\ref{H-1}) assuming that the solutions to the initial equation
(\ref{H}) are known. We seek for the intertwiner ${\cal L}$ in the
form of a linear, first-order differential operator
\begin{equation}
\label{L} {\cal L}=A(x) + B(x)d/dx,\end{equation}
 where the coefficients
$A$ and $B$ are to be determined. To this end we insert (\ref{L})
and the explicit form of the Hamiltonians ${\cal H}$ and
 $\widetilde{\cal H}$ into the intertwining relation (\ref{inter1})
and apply it to the solution $\phi$ of (\ref{H}). Assuming the
linear independence of derivative operators of different order
$d^k/dx^k,~k=0,1,2,3$, we obtain the following system of equations
for $A,~B$ and $\widetilde{v}$
\begin{eqnarray}\label{B1}
&&\frac{2}{q~m}~B{~'} ~=~ B~\left(\frac{1}{q~m}\right)^{\prime}~,
\\
&&\frac{2}{m}~A{'} + \frac{1}{m}~B{~''} +\left(\frac{1}{m}\right)'
B{~'}-B~q~\left[\frac{1}{q}\left(\frac{1}{m}\right)'\right]'  ~=~
B~\left({\widetilde{v}}-{v}\right),\label{A1}
\\
&& \frac{1}{m}~A{''} +
\left(\frac{1}{m}\right)'A{'}+B~q\left(\frac{v}{q}\right)^{\prime}~=A~({\widetilde{v}}-v)
\label{A2}~,\end{eqnarray} where the prime denotes differentiation
with respect to $x$ and arguments have been omitted. From (\ref{B1})
it follows
\begin{eqnarray}\label{B}
B = \frac{\beta}{\sqrt{q~m}}~,
\end{eqnarray}
where $\beta$  is an arbitrary constant of integration. Equations
(\ref{A1}) and (\ref{A2}) allow us to determine the potential
${\widetilde{v}}$ and the function $A$. Multiplying (\ref{A1}) by
$A$ and (\ref{A2}) by $B$ and subtracting the resulting equations we
obtain the equation with respect to $A$
\begin{eqnarray}
-\frac{A{''}}{q~m}+\left(\frac{2A{'}}{q~m~B}+
\frac{B~{''}}{q~m~B}+\frac{1}{q}\left(\frac{1}{m}\right)^{\prime}
\frac{B~{'}}{B}-\left[\frac{1}{q}\left(\frac{1}{m}\right)^{\prime}\right]^{\prime}
\right)A-B\left(\frac{v}{q}\right)^{\prime}-\frac{1}{q}\left(\frac{1}{m}\right)^{\prime}A{'}=0.
  \label{lasteq}
\end{eqnarray}
In order to solve this equation, we introduce a new auxiliary
function $K$ defined by $K=A/B$. Taking (\ref{B}) into account,
after simplification we arrive at the following nonlinear
 equation
$$\frac{d}{dx}\left[ \frac{1}{q~m}\left(-K^{'}+ K^2
\right)\right]-\frac{d}{dx}\left(\frac{v}{q}\right)
-\frac{d}{dx}\left[\frac{1}{q}\left(\frac{1}{m}\right)'~K\right]=0~,$$
from which the generalized Riccati equation follows
\begin{eqnarray}\label{ric}
\frac{1}{q~m}\left( -K^{'}+ K^2\right) -
\frac{v}{q}-\frac{1}{q}\left(\frac{1}{m}\right)'K=-\lambda,
\end{eqnarray}
where  $\lambda$ is an arbitrary integration constant. Equation
(\ref{ric})  turns into the  conventional  Riccati equation in
particular cases of $m=const$ and $q=const$. With $B$ from (\ref{B})
and $A=K/B$ the potential $\widetilde{v}$ can be expressed from
(\ref{A1}) in terms of $K$ and known potentials $v$, $m$ and $q$
\begin{eqnarray}
&& \widetilde{v}=v+2\sqrt{\frac{q}{m}} \frac{d}{dx}\frac{K}{\sqrt{
q~ m}} -\sqrt{\frac{q}{m}}
\frac{d}{dx}\left[\frac{1}{q}\frac{d}{dx}\left(\sqrt{\frac{q}{m}}~\right)
\right]\label{trV}.
\end{eqnarray}
The potential $\widetilde{v}$ will be finally determined after
finding the function $K$ from the Riccati equation (\ref{ric}), that
can be linearized and integrated by introducing an auxiliary
function ${\cal U}={\cal U}(x)$
\begin{eqnarray}\label{K1} K=-\frac{{\cal
U}^{~'}}{{\cal U}}.\end{eqnarray} Assuming that ${\cal U}$ is twice
continuously differentiable  and substituting (\ref{K1}) in
(\ref{ric}) we get the equation
\begin{equation}\label{Eq-U}
-\frac{1}{m}~{\cal U^{~''}}- \left(\frac{1}{m}\right)'{\cal
U}^{~'}+v ~{\cal U}=q~\lambda~{\cal U}~,\end{equation} which is
identical to the initial equation (\ref{H}) at  ${\cal E}=\lambda$.
Since the solutions of  (\ref{H}) with the given potentials $v$, $q$
and $m$ are known at all energies ${\cal E}$, hence we know the
solution ${\cal U}$. Once ${\cal U}$ is given, the function $K$ is
calculated via (\ref{K1}), which in turn determines $A$ by means of
$A=BK$ and (\ref{B}) $$ A= -\frac{1} {\sqrt{q~m}}\frac{{\cal
U}^{~'}}{{\cal U}}.$$ The constant of integration $\beta$ has been
taken to one. By insertion of $K$ in (\ref{trV}) we obtain the
explicit form of the transformed potential
\begin{eqnarray}
&&  \widetilde{v}=v-2\sqrt{\frac{q}{m}}
\frac{d}{dx}\left[\frac{1}{\sqrt{q~m}} \frac{{\cal U}^{~'}}{{\cal
U}} \right] - \sqrt{\frac{q}{m}}
\frac{d}{dx}\left[\frac{1}{q}\frac{d}{dx}\left(\sqrt{\frac{q}{m}}~\right)
\right],\label{tran-V}\end{eqnarray} Finally,  the  intertwiner
${\cal L}$ and the transformation solution $\widetilde{\phi}$ are
determined from (\ref{L}) and (\ref{inter2}), respectively
\begin{eqnarray}&& {\cal L}=\frac{1}{\sqrt{q~ m }}\left(\frac{d}{dx}+
K\right)=\frac{1}{\sqrt {q~ m}}\left[\frac{d}{dx} -\frac{{\cal
U}^{~'}}{{\cal U}} \right]~,  \label{L1}\\
&&\widetilde{\phi}~=~{\cal L}\phi = \frac{1}{\sqrt {q~
m}}\left[\frac{d}{dx} -\frac{{\cal U}^{~'}}{{\cal U}} \right]\phi~.
\label{w-phi}
\end{eqnarray}
 From (\ref{tran-V}) - (\ref{w-phi}) it
follows that the transformed potential $\widetilde{v}$, the
intertwiner ${\cal L}$ and solutions $\widetilde{\phi}$ depend not
only on the potential $v$ but also on the additional potentials $m$
and $q$. It is easy to check  that all expressions reduce to the
well known ones for the Schr\"odinger equation if potential
functions $m$ and $q$ are taken to be constant. When $q=const$ or
$m=const$ we get first-order Darboux transformations for
position-dependent mass Schr\"odinger equation \cite{suz-shul} or
for Schr\"odinger equation with weighted energy  \cite{Acta} as
particular cases of our approach.
\\
\\
{\sl Solutions at the energy of transformation.}
 Note
that relation (\ref{w-phi}) connects the solutions of two equations
(\ref{H}) and (\ref{H-1}) at an arbitrary energy except for
solutions at energy of transformation, ${\cal E}=\lambda$.
Evidently, at ${\cal E}=\lambda$ the action of Darboux
transformation (\ref{L1}) on the function  ${\cal U}$ and on
functions $\phi$ linearly dependent to ${\cal U}$ gives us ${\cal L}
{\cal U}=0$. In order to obtain a solution of the
transformed equation (\ref{H-1}) at  energy $\lambda$, we replace
${\cal U}$ by a linearly independent solution $\widehat{\cal U}$
\begin{equation}
\label{Liouv} \widehat{\cal U}= {\cal U}\int_{x_0}^x
dx'\frac{m(x')}{|{\cal U}(x')|^2}.
\end{equation}
The limits of integration depend on the boundary conditions.  The
direct substitution $\widehat{\cal U}$ into (\ref{Eq-U}) shows that
$\widehat{\cal U}$ really solves  the generalized Schr\"odinger
equation. The action of ${\cal L}$ on the function $\widehat{\cal
U}$ gives us a solution $\eta$ of the transformed equation
(\ref{H-1}) at energy $\lambda$
\begin{equation} \label{L-eta}
\eta={\cal L}\widehat{\cal U}=\sqrt{\frac{m}{q}}~\frac{1}{{\cal
U}}~.
\end{equation}
By using  the generalized Liouville formula (\ref{Liouv}) once
more, one can get a second solution $\widehat\eta$ of  (\ref{H-1})
at energy $\lambda$. For this ${\cal U}$ is replaced by $\eta$ in
(\ref{Liouv}) and with (\ref{L-eta}) we get
\begin{eqnarray}
\label{L-eta2} \widehat{\eta}= \eta\int_{x_0}^x dx'
\frac{m(x')}{|\eta(x')|^2} = \sqrt{\frac{m}{q}}~\frac{1}{{\cal
U}}\int_{x_0}^x dx'q(x') |{\cal U}(x')|^{2}.
\end{eqnarray}
Hence, the knowledge of all solutions of the initial equation
(\ref{H})  provides  the knowledge of all solutions of the
transformed equation (\ref{H-1}), including the solutions at energy
of transformation.
 Notice that if ${\cal U}$ replies to the bound state of
${\cal H}$, then the function $\eta$ defined by (\ref{L-eta}) at the
energy of transformation $\lambda$ cannot be normalized. Such bound
state is excluded from the spectrum of ${\cal H}$. Therefore, the
Hamiltonians ${\cal H}$ and $\widetilde{\cal H}$ are isospectral
except for the bound state with energy $\lambda$, which is removed
from the spectrum of ${\cal H}$.

\subsection{Supersymmetry}
  Now we consider the generalization of the supersymmetric formalism (SUSY) on the
generalized Schr\"odinger equation (\ref{1}) and show how one can
construct a Hamiltonian with an additional bound state by using
supersymmetry. As it is well  known, the SUSY algebra provides a
relation between superpartner Hamiltonians, which can be presented
in a factorized form in terms of Darboux transformation operators
${\cal L}$ and its adjoint ${\cal L}^\dag$.
 In our case
the scalar product of functions is defined by not the standard way
$(f,\chi)$ but with the weight of $q$: $(f,\chi)_q=\int
f^{*}(x)q(x)\chi(x)$. In this case instead of operator
$D^+=(CQ)^+=Q^+ C^+$ it is necessary to consider the operator
$D^\dag=q^{-1}D^+ q$. Therefore, the operator ${\cal L}^{\dag}$
adjoint to ${\cal L}=A + Bd/dx$ is determined as ${\cal
L}^\dag=q^{-1}\left(A-Bd/dx -dB/dx\right) q$. After simplification
we arrive at
\begin{eqnarray}\label{L-ad}
{\cal L}^{\dag} = \frac{1}{\sqrt{q~ m}}\left( - \frac{d}{dx} +K
\right) -\frac{1}{q}\frac{d}{dx}\sqrt{\frac{q}{m}}~
\end{eqnarray}
and the operator ${\cal L}^{\dag}$  satisfies  the intertwining
relation
\begin{equation}
\label{intert2} {\cal H}{\cal L}^{\dag}={\cal L}^{\dag}
\widetilde{\cal H}\,.
\end{equation}
The generalized  Schr\"odinger equations (\ref{H}) and (\ref{H-1})
can then be written as one single matrix equation in the form
\begin{eqnarray}
\left(
\begin{array}{cc}
{\cal H}-\lambda & 0 \\
0 & \widetilde{\cal H}-\lambda
\end{array}
\right) \left(
\begin{array}{c}
\phi \\
\tilde{\phi}
\end{array}
\right)=0. \label{mat}
\end{eqnarray}
On defining $H_s=$ diag$({\cal H},\widetilde{\cal H})$ and
$\Phi=(\phi,\tilde{\phi})^T$, the above matrix  Schr\"odinger
equation (\ref{mat}) can be written as
\begin{eqnarray}
[H_s-\lambda I]~\Phi = 0, \label{matrix}
\end{eqnarray}
where $I$ is the $2\times 2$ unity matrix.
 Similar to the case of the standard Schr\"odinger equation,  we
define two supercharge operators $Q$, $Q^\dag$ as follows:
\begin{eqnarray}
Q = \left(
\begin{array}{cc}
0 & 0 \\
{\cal L} & 0
\end{array}
\right)~,\qquad Q^\dag = \left(
\begin{array}{cc}
0 & {\cal L}^\dag \\
0 & 0
\end{array}
\right), \label{qqd}
\end{eqnarray}
where ${\cal L}$ and ${\cal L}^\dag$ are the operators given by
(\ref{L1}) and (\ref{L-ad}), respectively. One can show that the
matrix Hamiltonian $H_s$, as given in (\ref{matrix}), satisfies the
following conditions:
\begin{eqnarray}
 \label{susyalg1}
\left\{Q,Q \right\} &=& \left\{Q^\dag,Q^\dag \right\} = 0~, \vspace{.2cm} \\
\label{susyalg2} \left[Q,H_s \right] &=& \left[H_s,Q^\dag \right] =
0 \vspace{.2cm},
\end{eqnarray}
where $\{\cdot,\cdot\}$ and $[\cdot,\cdot]$ are the anticommutator
and the commutator, respectively. The relations (\ref{susyalg1}) are
trivially fulfilled, because the matrices in (\ref{qqd}) are
nilpotent. It is easily seen that equations (\ref{susyalg2}) are
equivalent to the intertwining ones (\ref{inter1}) and
(\ref{intert2}).

 Now, let us consider
the complementing relations of the supersymmetric algebra, that is,
the anticommutators $\left\{Q,Q^\dag \right\}$ and $\left\{Q^\dag,Q
\right\}$. For this, we calculate the operators ${\cal
L}^{\dag}{\cal L}$ and ${\cal L}{\cal L}^{\dag}$, and consider the
connections of them with our Hamiltonians ${\cal H}$ and
$\widetilde{\cal H}$. By using (\ref{L1}) and (\ref{L-ad}), we
arrive after some algebraic transformations at
\begin{eqnarray}
{\cal L}^{\dag}{\cal L}=-\frac{1}{q~m}
\partial_{xx}-\frac{1}{q}\left(\frac{1}{m}\right)^{\prime}
\partial_x+\frac{1}{q~m} \left(|K|^2-K^{\prime} \right)
  -\frac{1}{q}\left(\frac{1}{m}\right)^{\prime} K\,,~~~~~~~~~~~~~~&& \label{G-1} \\
{\cal L}{\cal L}^{\dag}=-\frac{1}{q~m}\partial_{xx}
-\frac{1}{q}\left(\frac{1}{m}\right)^{\prime}
\partial_x+\frac{1}{q~m}\left(|K|^2+K^{\prime} \right)
  +\frac{1}{m}\left(\frac{1}{q}\right)^{\prime} K
-\frac{1}{\sqrt{q~m}}\left[\frac{1}{q}\left(\sqrt{\frac{q}{m}}\right)^{\prime}
 \right]^{\prime}.&&\label{G-2}
\end{eqnarray}
We express the potential $v$ from the Riccati equation (\ref{ric})
in the form
\begin{eqnarray}\label{ric-V}
 v=\frac{1}{m}\left( -K^{\prime}+ K^2\right) -\left(\frac{1}{m}\right)^{\prime} K+q \lambda~.
\end{eqnarray}
Using this in (\ref{trV}), we get the following representation for
transformed potential
\begin{eqnarray}\label{v-K}
\tilde{v}=
 \frac{1}{m}(K^2+K^{\prime})
  +\frac{q}{m}\left(\frac{1}{q}\right)^{\prime}K-
 \frac{\sqrt{q}}{\sqrt{m}} \frac{d}{dx}\left[\frac{1}{q}\left(\sqrt{\frac{q}{m}}\right)^{\prime}
 \right]+q\lambda.
\end{eqnarray}
One can easily see that  the potential difference is determined as
\begin{eqnarray}\label{v-dif}
\widetilde{v}-v=2\sqrt{\frac{q}{m}} \frac{d}{dx}\frac{K}{\sqrt{ q~
m}} -\sqrt{\frac{q}{m}}
\frac{d}{dx}\left[\frac{1}{q}\frac{d}{dx}\left(\sqrt{\frac{q}{m}}~\right)
\right].
\end{eqnarray}
On further employing (\ref{ric-V}) and (\ref{v-K}), the formulae
(\ref{G-1}) and (\ref{G-2}) can be rewritten as
\begin{eqnarray}\label{factor}
{\cal L}^{\dag}{\cal L}=-\frac{1}{q}\left[\frac{\partial}{\partial
x}\left(\frac{1}{m}\right)\frac{\partial}{\partial x}\right]
+\frac{v}{q}-\lambda={\cal
H}-\lambda;\\
\label{factor1} {\cal L}{\cal L}^{\dag}=
-\frac{1}{q}\left[\frac{\partial}{\partial
x}\left(\frac{1}{m}\right)\frac{\partial}{\partial x}\right]
+\frac{\tilde{v}}{q}-\lambda=\widetilde{\cal H}-\lambda.
\end{eqnarray}
As can be easily seen, the anticommutation relation is
\begin{eqnarray}
\{Q,Q^{\dag} \}=
 \left(
\begin{array}{cc}
{\cal L}^{\dag}{\cal L}& 0 \\
0 & {\cal L}{\cal L}^{\dag}
\end{array}
\right)=\left(
\begin{array}{cc}
{\cal H}-\lambda & 0 \\
0 & \widetilde{\cal H}-\lambda
\end{array}
\right)={\cal H}_s-\lambda I. \label{mat-a}
\end{eqnarray}
In components, the latter equality reads
\begin{eqnarray}
\label{Hm0} &&{\cal H}={\cal L}^{\dag}{\cal L}+\lambda~,\\
\label{Hm1} &&\widetilde{\cal H}={\cal L}{\cal L}^{\dag}+\lambda~.
\end{eqnarray}
Note that as soon as the initial Hamiltonian ${\cal H}$ is presented
in the factorized form (\ref{Hm0}), one can get its supersymmetric
partner in a factorized form (\ref{Hm1}), too. Indeed, multiplying
equation (\ref{Hm0}) from the left by ${\cal L}$ and taking into
account the  intertwining relation (\ref{intert2}) we get
\begin{eqnarray}
{\cal L}~{\cal H}~\phi~=~{\cal L}~({\cal L}^\dag {\cal L}
+\lambda)~\phi~=~({\cal L}~{\cal L}^\dag +\lambda){\cal L}
~\phi=\widetilde{\cal H}~{\cal L}~\phi.
\label{H-new}
\end{eqnarray}
It means, that $\widetilde{\cal H}={\cal L}{\cal L}^{\dag}+\lambda$.

In summary, we obtained the explicit forms of the supersymmetric
partner Hamiltonians ${\cal H}$ and $\widetilde{\cal H}$. The
Hamiltonians (\ref{Hm0}) and (\ref{Hm1}) are compatible with their
definitions (\ref{H}) and (\ref{H-1}), respectively, if the
transformed potentials $v$ and $\tilde{v}$ are given by
(\ref{ric-V}) and (\ref{v-K}). Finally, taking the difference of the
factorized Hamiltonians  (\ref{Hm0}) and (\ref{Hm1}) gives the
potential difference (\ref{trV}) that we obtained for our Darboux
transformation. Hence, the Darboux transformation is equivalent to
the supersymmetry formalism.

The intertwining relation (\ref{intert2}) means that the operator
${\cal L}^{\dag}$ is also the transformation operator and realizes
the transformation from the solutions  of (\ref{H-1}) to the
solutions  of (\ref{H}). Evidently, one can interchange the role of
the initial generalized
 Schr\"odinger equation (\ref{H}) and its transformed
counterpart (\ref{H-1}). To this end, let us express the operators
${\cal L}$ and ${\cal L}^{\dag}$ in terms of functions $\eta$ given
in (\ref{L-eta}), which are solutions to the equation (\ref{H-1}) at
the energy of transformation $\lambda$. First, we rewrite $K$ by
using ${\cal U}$ from the relation (\ref{L-eta})
\begin{displaymath}
K=-\frac{{\cal U}~'}{{\cal
U}}=\frac{1}{2}\frac{q^{'}}{q}-\frac{1}{2}\frac{m^{'}}{m}+\frac{\eta'}{\eta}.
\end{displaymath}
 Using this in (\ref{L1}) and (\ref{L-ad}), we obtain after
 simplifications
\begin{equation}
\label{L1-eta}{\cal L}^{\dag}= \frac{1}{\sqrt {q~ m}}
\left(-\frac{d}{dx} +\frac{\eta'}{\eta}\right)~, ~~~~ {\cal L} =
\frac{1}{\sqrt{q~ m}}\left( \frac{d}{dx} +\frac{\eta{'}}{\eta}
\right) +\frac{1}{q}\frac{d}{dx}\sqrt{\frac{q}{m}}
\end{equation}
Obviously, the function $\eta$ is also a transformation function.
Notice, ${\cal L}^{\dag}\eta=0$, meaning that $\eta$ belongs to the
kernel of the operator ${\cal L}^{\dag}$. As one can see from
(\ref{L1-eta}) and (\ref{L-eta2}), the application of the operator
${\cal L}^{\dag}$ to the second linearly independent solution $\hat
\eta$  of equation (\ref{H-1}) gives back the solutions ${\cal U}$
of the initial problem at the energy of transformation. Indeed,
\begin{eqnarray}
{\cal L}^{\dag}\hat \eta=\frac{1}{\sqrt {q~ m}} \left(-\frac{d}{dx}
+\frac{\eta'}{\eta}\right) ~\eta
 \int_{x_0}^x dx'\frac{m(x')}{|\eta(x')|^2}
=-\sqrt{\frac{m}{q}}~\frac{ {1}}{\eta}=-{\cal U}. \nonumber
\end{eqnarray}
Hence, the solution at the energy of transformation $\lambda$ takes
the form
\begin{eqnarray}
\label{tr-U}{\cal U}=\sqrt{\frac{m}{q}}~\frac{{1}}{\eta}
\end{eqnarray}
and, in principle, can reply to the new bound state. The second
linearly independent solution $\widehat{\cal U}$ of (\ref{H}) at
energy $\lambda$ can be written in terms of $\eta$ as follows:
\begin{eqnarray}
\label{tr-U2}\widehat{\cal U}=\sqrt{\frac{m}{q}}\frac{
1}{\eta}\int^{x}dx'q|\eta|^2.
\end{eqnarray}
Introducing the function $\widetilde{K}=\widetilde{K}(x)$ by
$\widetilde{K}=\eta'/\eta$ and taking into account
$\frac{1}{\sqrt{q~ m}}
\left(\frac{1}{2}\frac{q^{'}}{q}-\frac{1}{2}\frac{m^{'}}{m}\right)=
\frac{1}{q}\frac{d}{dx}\sqrt{\frac{q}{m}}$, the expressions
(\ref{L1-eta}) for operators ${\cal L}^{\dag}$ and ${\cal L}$ can be
rewritten as
\begin{equation}
{\cal L}^{\dag}= \frac{1}{\sqrt {q~ m}} \left(-\frac{d}{dx}
+\widetilde{K} \right), ~~~~
 {\cal L} = \frac{1}{\sqrt{q~ m}}\left( \frac{d}{dx}
+\widetilde{K} \right) +\frac{1}{q}\frac{d}{dx}\sqrt{\frac{q}{m}}.
\end{equation}
Using (\ref{v-dif}) the potential $v$
 can be expressed in terms of
$\widetilde{v}$ as follows:
\begin{eqnarray}\label{tr-wV}
v={\widetilde v}-\sqrt {\frac{q}{m}}\left[2\frac{d}{dx}\left(
\frac{\widetilde{K}}{\sqrt{q~m}}\right)+
\frac{d}{dx}\left(\frac{1}{q}\frac{d}{dx}\left(\sqrt{\frac{q}{m}}~\right)
\right)\right]~, ~~~~\widetilde{K}=\frac{\eta'}{\eta}
\end{eqnarray}
 and corresponding solution $\phi$ are given by
 \begin{eqnarray}
\label{tr-wphi} \phi={\cal L}^{\dag}~{\widetilde\phi} =
 \frac{1}{\sqrt{q~ m}}\left[-\frac{d}{dx} + \frac{\eta'}{\eta} \right]{\widetilde
\phi}.
\end{eqnarray}
Thus, the function $\eta$ becomes a transformation function for the
operator ${\cal L}^{\dag}$, which performs the transformation from
the potential $\widetilde{v}$ to the potential $v$ and from the
solutions of (\ref{H-1})  to the solutions of (\ref{H}). If within
the first procedure (\ref{L1})--(\ref{w-phi}) we constructed the
potential $\widetilde{v}$ with one bound state removed, now we can
construct the potential $v$ with an additional bound state. Note
that we have established  a one-to-one correspondence between the
spaces of solutions of equations (\ref{H}) and (\ref{H-1}). The
operators ${\cal L}$ and ${\cal L}^{\dag}$ realize  this
correspondence for any ${\cal E}\ne\lambda$. If ${\cal E}=\lambda$,
the correspondence is ensured by mapping $\eta\longleftrightarrow
\widehat{\cal U}$ and $\widehat{\eta}\longleftrightarrow \cal{U}$.

In particular cases our generalized Darboux transformations are
reduced correctly to the known expressions. In the case with a
constant weighted energy potential, e.g. $q(x)=1$, from our
supersymmetry approach we get the supersymmetry
 for effective mass Schr\"odinger equation \cite{Milanov,suz-shul}.
In the case with constant mass $m(x)=m_0$ from our approach  we
obtain the supersymmetry for Schr\"odinger equation with weighted
energy \cite{Acta}. Finally, if $m(x)=m_0$ and $q(x)=1$, our
expressions of supersymmetric algebra are correctly reduced to the
conventional ones for the standard Schr\"odinger equation (see, e.g.
\cite{matveev}).

\section{Higher order  Darboux
transformations}

 In this section by considering iterative
applications of first-order Darboux transformations $n$ times we
obtain the $n$th order Darboux transformation and show that  the
$n$th order Darboux transformation  can be expressed in terms of
solutions of an initial equation, with no use of the solutions to
intermediate equations.

\subsection{Chain of  Darboux transformations}
Now we consider iteration of one-step transformations  we have
obtained in the previous section, in order to construct higher order
Darboux transformations. To this end we show that the intertwining
operator ${\cal L}$ of the $n$th order
 can
be obtained from a sequence of $n$ first-order Darboux
transformations, like conventional Schr\"odinger equation, and
create a chain of exactly solvable Hamiltonians ${\cal H}_1, {\cal
H}_2,..., {\cal H}_n$. Let us define the second-order Darboux
transformation as a sequence of two first-order Darboux
transformations ${\cal L}_1$ and ${\cal L}_2$
 \begin{equation}
 \label{SL}
 {\cal L}={\cal L}_2{\cal L}_1~,
\end{equation}
where   ${\cal L}_1$  is actually ${\cal L}$ given in (\ref{L1}).
Let ${\cal U}_1$ be an auxiliary solution of (\ref{H}) at energy
$\lambda_1$. Then we have
\begin{equation}
\label{SL1} {\cal L}_1=\frac{1}{\sqrt {q~ m}}\left(\frac{d}{dx}+
K_1\right),~~~ K_1=-\frac{ {\cal U}^{~'}_{1}}{{\cal U}_1}~.
\end{equation}
The operator ${\cal L}_2$ is determined as follows:
\begin{equation}
 \label{SL2}
 {\cal L}_2=\frac{1}{\sqrt {q~ m}}\left(\frac{d}{dx}+ K_2\right),~~~
 K_2=-\frac{\chi_1{'}}{\chi_1}~.
\end{equation}
The function $\chi_1$ is obtained by means of the first-order
Darboux transformation (\ref{SL1}), applied to an auxiliary solution
${\cal U}_{2}$ of equation (\ref{H}) at energy $\lambda_2$
\begin{equation}
 \label{chi}
 \chi_1={\cal L}_1{\cal U}_2=\frac{1}{\sqrt {q~ m}}\left(\frac{d}{dx}+K_1
\right){\cal U}_2~.
\end{equation}
Clearly, $\chi_1$ is the solution of the transformed equation
(\ref{H-1}) with the potential  $v_1=\widetilde{v}$, defined as in
(\ref{trV}), and $\chi_1$ can be taken as a new transformation
function for the Hamiltonian ${\cal H}_1=\widetilde{\cal H} $ to
generate a new potential
\begin{equation}
 \label{V2}
v_2=v_1+2\sqrt{\frac{q}{m}} \frac{d}{dx}\frac{K_2}{\sqrt{q~ m}}
-\sqrt{\frac{q}{m}}
\frac{d}{dx}\left[\frac{1}{q}\frac{d}{dx}\left(\sqrt{\frac{q}{m}}~\right)
\right],
\end{equation}
 and
corresponding solutions
\begin{equation}\label{Sphi2}
\phi_2={\cal L}_2\phi_1=\frac{1}{\sqrt {q~m}}
\left(\frac{d}{dx}+K_2\right)\phi_{1},~~\phi_{1}={\cal L}_1\phi.
\end{equation}
The function  $\phi_{1}=\widetilde{\phi}$, defined as in
(\ref{w-phi}),  is the solution of equation (\ref{H-1}) with the
Hamiltonian ${\cal H}_1=\widetilde{\cal H}$. In summary, the action
of the 2nd-order operator (\ref{SL}) on solutions $\phi$ of the
generalized equation (\ref{H}) leads to solutions $\phi_2$ of
\begin{equation} \label{H2-eq}
  {\cal H}_2~\phi_2(x)={\cal E}\phi_2(x),\qquad
 {\cal H}_2=
  -  \frac{1}{q}\left[\frac{d}{dx}\left(\frac{1}{m }\right)\frac{d}{dx}\right]+
 \frac{v_2}{q},\end{equation} given by
\begin{equation}\label{L-phi}
\phi_2={\cal L}~\phi={\cal L}_2~{\cal L}_1~\phi.
\end{equation}

Iterating the procedure $n$ times in regard to the given operator
${\cal H}$ leads to the operator ${\cal H}_n$, which satisfies the
intertwining relation
\begin{equation}\label{intert-n}
{\cal L}{\cal H}={\cal H}_n{\cal L}.
 \end{equation}
The transformed potentials $ v_n $ satisfy the following recursion
relation
\begin{equation}\label{VM-pr}
 v_n=v_{n-1}+2 \sqrt{\frac{q}{m}}\frac{d}{dx}\frac{K_{n}}{\sqrt{q~
m}} -\sqrt{\frac{q}{m}}
\frac{d}{dx}\left[\frac{1}{q}\frac{d}{dx}\left(\sqrt{\frac{q}{m}}~\right)
\right],
 \end{equation}
the corresponding solutions are
 \begin{equation}\label{phi-m}
 \phi_n={\cal L}\phi={\cal L}_n\phi_{n-1}={\cal L}_n{\cal L}_{n-1}...{\cal L}_1\phi,
\end{equation}
where ${\cal L}$ is the $n$th order operator:
\begin{equation}\label{LM}
{\cal L}={\cal L}_n{\cal L}_{n-1}...{\cal L}_1,~~{\cal L}_n
=\frac{1}{\sqrt {q ~m}}\left(\frac{d}{dx}+K_n\right),
~~K_n=-\frac{\chi_{n-1}^{'}}{\chi_{n-1}}.
\end{equation}
Thus,  the chain of $n$ first-order Darboux transformations results
in a chain of exactly solvable Hamiltonians ${\cal H}\to {\cal
H}_1\to...\to{\cal H}_n$. When ${\cal L}$ is the $n$th order
differential operator and the intertwining relation (\ref{intert-n})
is valid,  the so-called $n$th order supersymmetry arises, like for
the ordinary Schr\"odinger equation \cite{Andrianov}.

\subsection{Darboux
transformation of the $n$th order} Now we  show that  iteration
procedure of the $1$st-order Darboux transformations in
(\ref{VM-pr}) - (\ref{LM}) can be removed and transformed potentials
$ v_n$ and solutions $ \phi_n$ can be expressed in terms of the
initial potentials $v$, $m$ and $q$ and the family $\{{\cal U}_j\}$,
$j=1,2,...,n$ of auxiliary  solutions of the initial equation
(\ref{H}) at energies $\lambda_j$. Consider now the 2nd order
transformation in  detail. Using the explicit expression for $v_1$
which appears in the first-order Darboux transformation (\ref{trV}),
we present the formula (\ref{V2}) for the potential $v_2$ as
\begin{equation}\label{V2K}
v_2=v +2 \sqrt{\frac{q}{m}}\frac{d}{dx}\frac{K}{\sqrt{q~ m}} -2
\sqrt{\frac{q}{m}}
\frac{d}{dx}\left[\frac{1}{q}\frac{d}{dx}\left(\sqrt{\frac{q}{m}}~\right)
\right]~,~~~~K=K_1+K_2~.
\end{equation}
In order to find $K$  transform $K_2=-\chi_1{'}/\chi_1$,
representing $\chi_1$ as
\begin{equation} \label{chi1}
\chi_1=\frac{1}{\sqrt {q~ m}}\frac{W_{1,2}}{{\cal U}_1}~,
\end{equation}
where $W_{1,2}={\cal U}_1{\cal U}_2{'}-{\cal U}_1{'}{\cal U}_2$ is
 the Wronskian of the functions ${\cal U}_1$ and ${\cal U}_2$.
 Substituting (\ref{chi1}) into the formula (\ref{SL2}) for $K_2$,
 we get
\begin{equation}\label{SK-2}
  K_2=-\frac{\chi_1{'}}{\chi_1}= -\frac{d}{dx}\left[\ln \left(\frac{1}{\sqrt {q~ m}}\frac{W_{1,2}}{{\cal
  U}_1}\right)\right]
  \end{equation}
and with account $K_1=-{\cal U}_1{'}/{\cal U}_1$ we obtain
 \begin{equation}\label{K-12}
K=-\frac{d}{dx}\left[\ln \frac{W_{1,2}}{\sqrt{q~ m}}\right].
   \end{equation}
With the last expression after some manipulations, the new potential
$v_2$ can be expressed as
\begin{eqnarray}\label{Sv-2}
v_2=v-2\frac{\sqrt{q}}{\sqrt{m}}\frac{d}{dx}\left[ \frac{1}{W_{1,2}}
~\frac{d}{dx}\frac{W_{1,2}}{\sqrt{q~m}}\right]- 2~\sqrt{\frac{q}{m}}
\frac{d}{dx}\left[\frac{1}{q}\frac{d}{dx}\left(\sqrt{\frac{q}{m}}~\right)~\right]
\end{eqnarray}
and finally in the form
\begin{equation}\label{V2tr}
v_2=v -2 \sqrt{\frac{q}{m}}\frac{d}{dx}\left[
\frac{\sqrt{m}~\frac{d}{dx}\left(W_{1,2}/m\right)}{\sqrt{q}~W_{1,2}}
 \right]
.\end{equation} Find now the corresponding functions $\phi_2$. To
this end let us transform the relation (\ref{Sphi2}). By analogy
with $\chi_1$ the function   $\phi_1$ can be written in terms of the
Wronskian  $W_{1,{\cal E}}={\cal U}_1\phi{'}-{\cal U}_1{'}\phi$:
\begin{equation} \label{phi1}
\phi_1=\frac{1}{\sqrt {q~ m}}\frac{W_{1,{\cal E}}}{{\cal U}_1}~,
\end{equation}
Let us now calculate the derivative of $\phi_1$
$$ \phi_1^{'}=({\cal L}_1\phi){'}=
\left(\frac{1}{\sqrt{q~m}~{\cal U}_1}\right)'W_{1,{\cal E}}
 + \frac{1}{\sqrt{q~m}}\phi{''}-
\frac{1}{\sqrt{q~m}}\frac{{\cal U}_1{''}}{{\cal U}_1}\phi.
$$
Making use of the last expression and the relation (\ref{SK-2}) for
$K_2$ in (\ref{Sphi2}), we obtain, after some simplification, the
solutions as follows
\begin{eqnarray}\label{S-phi-2}
\phi_2=\frac{1}{q~m }\left(\phi{''}-\frac{{\cal U}_1{''}}{{\cal
U}_1}\phi\right) -\frac{1}{q~m } \frac{ W_{1,2}'}{ W_{1,2}}
\frac{W_{1,{\cal E}}}{{\cal U}_1}~=\frac{1}{q~m }\frac{W_{1,2,{\cal
E}}}{W_{1,2}}~.
\end{eqnarray}
It is easily seen from (\ref{V2tr}) and (\ref{S-phi-2}) that due to
the 2nd order Darboux transformations, the potential and solutions
are completely expressed in terms of the known potential functions,
$v, m$ and $q$ and the solutions ${\cal U}_1, {\cal U}_2, \phi({\cal
E})$ of the initial equation, with no use of the solutions to the
intermediate equation with the potential $v_1(x)$.

Clearly,  for the next transformation step to be made, one should
take  a new transformation function $\chi_2$, that corresponds to
the potential $v_2$ at energy $\lambda_3$. The solution $\chi_2$ can
be obtained by applying the operator ${\cal L}={\cal L}_2{\cal L}_1$
to the transformation solution ${\cal U}_3$ at energy $\lambda_3$,
that is
\begin{eqnarray}
\chi_2={\cal L}_2{\cal L}_1{\cal U}_3. \nonumber
\end{eqnarray}
According to (\ref{S-phi-2}), the solution  $\chi_2$ can be written
as
\begin{eqnarray}\label{chi-2}
\chi_2=\frac{1}{q~m}\frac{W_{123}}{W_{12}}
\end{eqnarray}
and can be used to produce a new transformed operator ${\cal L}_3$,
given by
\begin{eqnarray}
{\cal L}_3=\frac{1}{\sqrt{q ~m}}\left(\frac{d}{dx}+K_3
\right),~~~K_3=-\frac{\chi_{2}{'}}{\chi_{2}} \nonumber
\end{eqnarray}
for generating a new potential $v_3$ with corresponding solutions
$\phi_3$ and  so on according to (\ref{VM-pr}) - (\ref{LM}). In this
way we can express the transformed potentials $v_n$ of any order in
terms of the initial potentials $v$, $q$, the effective mass $m$ and
the family of auxiliary solutions ${\cal U}_{j}$, $j=1,2,,...n$ of
the initial equation (\ref{H}) at energies $\lambda_j$, which
different from each other:
\begin{equation}\label{Sv-n}
v_n=v +2 \sqrt{\frac{q}{m}}\frac{d}{dx}\frac{K}{\sqrt{q~ m}} -
n~\sqrt{\frac{q}{m}}
\frac{d}{dx}\left[\frac{1}{q}\frac{d}{dx}\left(\sqrt{\frac{q}{m}}~\right)
\right]~,~~~~K=K_1+K_2+...+K_n~.
\end{equation}
In the case of $n$-order transformation over the initial potential
$v$ equation (\ref{Sv-n}) gives us the $n$-SUSY partner potential
$v_n$.  This construction enables us to generate a family of new
Hamiltonians of any order and corresponding solutions directly from
the initial Hamiltonian and solutions without generating
intermediate Hamiltonians. In this section we constructed $n$th
order Darboux transformations by using a chain of first-order
Darboux transformations and shown that the $n$th order Darboux
transformations are equivalent to the resulting action of the
iterative method.

\section{The integral form of Darboux transformations.}
In this section the  generalized Darboux transformations are
represented  in the integral form and applied to construction of
Hamiltonians with the same spectrum as the initial one (totally
isospectral Hamiltonians), and differing by one bound state and by
two bound states.

\subsection{First- and second-order integral transformations}

At the beginning  consider {\sl the first-order transformation}.
Multiplying both sides of  equation (\ref{H}) for $\phi$ by ${\cal
U}_1$  and subtracting from the obtained expression the equation
(\ref{H}) for ${\cal U}_1$ multiplied by $\phi$, we arrive at
\begin{equation}
\label{W-1E} \frac{d}{dx}\left(\frac{W_{1,{\cal
E}}}{m}\right)=(\lambda_1-{\cal E})~q~{\cal U}_1~\phi~.
\end{equation}
The last expression can be easily integrated:
\begin{eqnarray}
\label{wron-in} W_{1,{\cal E}}= m~\left((\lambda_1-{\cal E})\int^{x}
q(x'){\cal U}_1(x')\phi(x')dx'+ C\right)~,
\end{eqnarray}
where $C$ is a constant of integration. Inserting the integration
result into the formula (\ref{phi1}) for $\phi_1$, we arrive at the
integral form for the $1$st order transformed solutions:
 \begin{eqnarray}
\label{Phi1-int} \phi_1=  \sqrt{\frac{m}{q}}\frac{1}{{\cal
U}_1}\left( C+(\lambda_1-{\cal E})\int^{x}q(x')
 {\cal U}_1(x')\phi(x')dx'\right)~.
\end{eqnarray}
The auxiliary solutions $\chi_{1}$, taking at the energy ${\cal
E}=\lambda_2$, can be written as
 \begin{eqnarray}
\label{chi1-int} \chi_1=  \sqrt{\frac{m}{q}}\frac{1}{{\cal
U}_1}\left( C_1+(\lambda_1-\lambda_2)\int^{x}q(x')
 {\cal U}_1(x'){\cal U}_2(x')dx'\right)~,
\end{eqnarray}
where it was used $ W_{1,2}=
m\left((\lambda_1-\lambda_2)\int^{x}q~{\cal U}_1{\cal U}_2~dx'+
C_1\right)$. The integration limits depend on the boundary
conditions.

Now  consider {\sl the second-order transformation}. By analogy,
using
$$W_{\chi_1,\phi_1}= m~\left((\lambda_2-{\cal E})\int^{x}
q(x')\chi_1(x')\phi_1(x')dx'+ C\right)$$ in (\ref{Sphi2}),  we shall
get the integral form for the 2nd order transformed solutions
presented in terms of the 1st order solutions $\phi_1$ and
$\chi_{1}$
\begin{eqnarray}
\label{phi2-int}\phi_2=\frac{1}{\sqrt {q~
m}}\frac{W_{\chi_1,\phi_1}}{\chi_1}=
\sqrt{\frac{m}{q}}\frac{1}{\chi_1}\left(C+(\lambda_2-{\cal E})
\int^{x}q(x'){\chi}_1(x')\phi_1(x')dx'\right).
\end{eqnarray}
Evidently,  the  solutions $\phi_2$ can be expressed directly in
terms of the solutions to the initial equation (\ref{H}). For this
transform the expression (\ref{S-phi-2}) for $\phi_2$ having regard
to equation (\ref{H}) for $\phi$ and ${\cal U}_1$, using
(\ref{wron-in})  and $\frac{1}{m}[W_{1,2}]'=\left[\frac{ W_{1,2}}{
m}\right]'-\left[\frac{1}{m}\right]'W_{1,2}$ we obtain
 \begin{eqnarray}\label{Phi2-int}
 &&\phi_2
 = (\lambda_1- {\cal E} )~ \phi -
 \frac{
{\cal U}_2 (\lambda_1-\lambda_2) W_{1,{\cal E}}}{
m[C_1+(\lambda_1-\lambda_2)\int^{x}q(x')
{\cal U}_1(x'){\cal U}_2(x')dx']} =\nonumber\\
&&
 = (\lambda_1- {\cal E} )~ \phi~ -
 \frac{ {\cal U}_2
\left[C+(\lambda_1- {\cal E} )
\int^{x}q(x')
 {\cal U}_1(x')\phi(x')dx'\right]
 }
 {c_1+\int^{x}q(x')
 {\cal U}_1(x'){\cal U}_2(x')dx'}~,
\end{eqnarray}
where we have introduced $c_1=C_1/(\lambda_1-\lambda_2)$.
 Using the integral presentation of Wronskian in
(\ref{V2tr}), the transformed potential $v_2$ can be written as
\begin{eqnarray}
\label{V2-int}  v_2=
v-2\sqrt{\frac{q}{m}}\frac{d}{dx}\left(\frac{1}{\sqrt{q m }}~
\frac{q~{\cal U}_2{\cal U}_1}{c_1+ \int^{x}dx'~q(x'){\cal
U}_1(x'){\cal U}_2(x')} \right)~.
\end{eqnarray}
Thus, we get the integral form of the first- and second-order
Darboux transformations for the potentials and solutions. Note, the
spectrum of Hamiltonian ${\cal H}_2$ with the potential $v_2$
differs from the spectrum of the initial Hamiltonian  ${\cal H}$ by
two bound states $\lambda_1$ and $\lambda_2$.

The formulae (\ref{V2-int}) and (\ref{Phi2-int}) can be rewritten in
a more simple form
\begin{eqnarray}
\label{V2-K}  v_2(x)=
v(x)-2\sqrt{\frac{q(x)}{m(x)}}\frac{d}{dx}\left(\frac{1}{\sqrt{q(x)
m(x)}}~K(x,x)\right)~,
\end{eqnarray}
\begin{eqnarray}
\label{phi2-K} \phi_2=\phi- \int^{x}K(x,x') \phi(x')dx'
\end{eqnarray}
with  the operator kernel $K(x,x')$ determined as
\begin{eqnarray}\label{K-deg}
K(x,x')=\frac{{\cal U}_2(x)q(x')~{\cal U}_1(x')}{c_1+
\int^{x}dx'~q(x'){\cal U}_1(x'){\cal U}_2(x')}.
\end{eqnarray}
In (\ref{phi2-K}) the constant $C$, connected with the Wronskian
$W_{1,{\cal E}}$, is chosen to be zero. It is interesting to note,
that at $q(x)=const$ and $m(x)=const$ the formulae (\ref{V2-K}) and
(\ref{phi2-K}) look like the inverse problem ones for potentials and
solutions \cite{Levitan,Marchenko} obtained with the degenerate
kernels $K(x,x')$ except  for the form of transformation operator
(\ref{K-deg}). The operator kernel (\ref{K-deg}) differs from the
inverse problem kernel $K(x,x')$ not only  by $q(x)\ne const$ but by
auxiliary functions ${\cal U}_1(x)$ and ${\cal U}_2(x)$, which
correspond different energies $\lambda_1\ne\lambda_2$. In the next
section we shall consider the case when the 2nd order Darboux
transformations
 allow to change  the spectrum of a given Hamiltonian on one bound
 state that corresponds to ${\cal U}_1(x)={\cal
U}_2(x)$.

\subsection{Hamiltonians ${\cal H}_2$ differing by one bound state and
completely isospectral Hamiltonians} The first-order supersymmetry,
above considered in the differential and integral forms, gives us
opportunities to construct isospectral Hamiltonians differing by one
bound state. It is interesting to note,  if we use Darboux
transformation in its integral form, then we directly from
(\ref{Phi1-int}) obtain the solution of the partner  equation
(\ref{H-1})
 at energy of transformation ${\cal
E}=\lambda_1$ which with an accuracy of an arbitrary constant
coincides with (\ref{L-eta})
$$\eta=\sqrt{\frac{m}{q}}~\frac{C}{{\cal U}_1}~.$$

Now we show how using double Darboux transformations to generate
Hamiltonians ${\cal H}_2$, the spectrum of which differ from the
spectrum of the initial Hamiltonian ${\cal H}$ by one bound state
and how to construct a family of isospectral Hamiltonians ${\cal
H}_2$, the spectrum of which completely coincide with the spectrum
of ${\cal H}$. For the $2$-nd order transformation we used the
transformation function $\chi_1(x)$ obtained within the first step,
 $\chi_1={\cal L}_1{\cal U}_2=\frac{1}{\sqrt {q~ m}}\left(\frac{d}{dx}
-\frac{{\cal U}_1{'}} {{\cal U}_1} \right){\cal U}_2$ with
$\lambda_1\ne\lambda_2$. But for $\lambda_2=\lambda_1$ we have
${\cal U}_2={\cal U}_1$ and it leads to $\chi_1={\cal L}_1{\cal
U}_2=0$.
 For the second transformation one  can use the function $\chi_1$
constructed by means of a linear combination of the solutions $\eta$
and $\hat{\eta}$: $\chi_1=c_1\eta+c_2\widehat{\eta}$. For our aim we
take a linear combination as follows
\begin{eqnarray}\label{chi1-int}
\chi_1=c_1 \eta+\widehat{\eta}= \sqrt{\frac{m}{q}}\frac{1}{{\cal
U}_1}\left(c_1+ \int^{x}dx'q(x'){\cal U}_1^2(x')\right),
\end{eqnarray}
By analogy with (\ref{K-12}) we  calculate $K=K_1+K_2$ with
$K_1=-{\cal U}_1{'}/{\cal U}_1$ and $K_2=-\chi_1{'}/\chi_1$. After
simplification we have
$$
K_2=\frac{{\cal U}_1{'}}{{\cal
U}_1}-\frac{\frac{d}{dx}\left[\sqrt{\frac{m}{q}}\left(c_1+\int^{x}dx'q~{\cal
U}_{1}^{~2}\right)\right]}{\sqrt{\frac{m}{q}}
\left(c_1+\int^{x}dx'q~{\cal U}_{1}^{2}\right)}~
$$
and
\begin{eqnarray}\label{K-int}K(x)=-\frac{d}{dx}\left(\ln\sqrt{\frac{m}{q}}~\right)-
\frac{q~{\cal U}_1^{2}}{\left(c_1+\int^{x}dx'q~{\cal U}_1^{2}\right)
}~.
\end{eqnarray}
Plugging the last expression into the formula (\ref{V2K}) which
defines the potential, after some transformations we arrive at
\begin{eqnarray}
\label{V-int_2}  v_2=
v(x)-2\sqrt{\frac{q}{m}}\frac{d}{dx}\left(\sqrt{\frac{q}{m}}~
\frac{{\cal U}_1^{2}}{c_1+ \int^{x}dx'~q(x'){\cal U}_1^{2}(x')}
\right)~.
\end{eqnarray}
 With
$\chi_1$ defined by (\ref{chi1-int}) and $\phi_1$ represented by its
integral form (\ref{Phi1-int}) the relation (\ref{phi2-int}) leads
to
 \begin{eqnarray}
\label{Phi2-tr1} \phi_2= \phi(\lambda_1-{\cal E}) -\frac{{\cal
U}_1}{c_1 +\int^{x}q(x')
 {\cal U}_1^{2}(x')dx'}\left( C+ (\lambda_1-{\cal E})\int^{x}q(x')
 {\cal U}_1(x')\phi(x')dx'\right) .
\end{eqnarray}
It should be noted, that in difference from the differential
approach, the relations (\ref{V-int_2}) and (\ref{Phi2-tr1}) for the
new potential $v_2$ and the solution $\phi_2$ can be obtained
directly from (\ref{V2-int}) and (\ref{Phi2-int}), which replies the
transformations with two bound states, if one takes ${\cal U}_1
={\cal U}_2$.

It is worth noting that the auxiliary function $\chi_1$ can be
determined as follows
\begin{eqnarray}\label{chi1-int1}
\chi_1=\eta+\Gamma\hat{\eta}= \sqrt{\frac{m}{q}}\frac{1}{{\cal
U}_1}\left(1+ \Gamma\int^{x}dx'q(x'){\cal U}_1^2(x')\right)
\end{eqnarray}
Then the
potential  $v_2$ and solutions $\phi_2$ are rewritten in the form
\begin{eqnarray}
\label{V_phase}  v_2=
v-2\sqrt{\frac{q}{m}}\frac{d}{dx}\left(\sqrt{\frac{q}{m}}~
\frac{~\Gamma{\cal U}_1^{2}}{1+ \Gamma\int^{x}dx'~q(x'){\cal
U}_1^{2}(x')} \right)~,
\end{eqnarray}
 \begin{eqnarray}
\label{Phi2-phase} \phi_2= (\lambda_1-{\cal E})\phi
-\frac{{\Gamma\cal U}_1}{1+\Gamma\int^{x}q(x')
 {\cal U}_1^{2}(x')dx'} \left[C+(\lambda_1-{\cal
E})\int^{x}q(x')
 {\cal U}_1(x')\phi(x')dx'\right]~.
\end{eqnarray}
 In principle,  the formulae
(\ref{chi1-int1}) - (\ref{Phi2-phase}) coincide with
(\ref{chi1-int}), (\ref{V-int_2}), (\ref{Phi2-tr1}) at
$\Gamma=1/c_1$, but they are more suitable for physical
applications. Here, the constant  $\Gamma$  plays a role of a
normalization constant of the new bound state $\lambda_1$ provided
the other spectral characteristics coincide.

By analogy with the differential Darboux transformations, the
solution of the generalized equation (\ref{1}) with the potential
(\ref{V_phase}) at  energy of transformation $ {\cal E}=\lambda_1 $
can be achieved  by means of operator ${\cal L}_2$ acting on the
solution $\eta$ from (\ref{L-eta}), obtained within the first
transformation step
 \begin{eqnarray}\label{eta2-tr}\eta_2 ={\cal
L}_2\eta=\frac{1}{\sqrt {q~
m}}\left(\frac{d}{dx}-\frac{\chi'_1}{\chi_1}\right)\sqrt{\frac{m}{q}}
\frac{1}{{\cal U}_1}~.
\end{eqnarray}
After transformations (\ref{eta2-tr}) with an account of
(\ref{chi1-int1}) we get
\begin{equation}\label{eta2}
\eta_2 =-\frac{\Gamma{\cal U}_1}{1+\Gamma\int^{x}dx'q(x'){\cal
U}_1^2(x')}.
\end{equation}
 Note, the solution $\eta_2 $ can be directly  obtained from (\ref{Phi2-phase})
 at ${\cal E}=\lambda_1$. Obtaining solutions at energies of transformation from the general
 formulae at arbitrary energies is one of advantages of integral
 transformations.
 The relations
for potential (\ref{V_phase}) and solutions (\ref{Phi2-phase}) can
be rewritten in the form (\ref{V2-K}) and (\ref{phi2-K})
 with the operator kernel $K(x,x')$ determined  as
$$K(x,x')=\frac{\Gamma{\cal U}_1(x)q(x')~{\cal U}_1(x')}{1+ \Gamma\int^{x}dx'~q(x'){\cal
U}_1^2(x')}~.$$ At $m(x)=const=m_0$ and $q(x)=const$
 our generalized expressions
 are correctly reduced to the integral equations of inverse problem for the
standard Schr\"odinger equation with the degenerate kernel of
transformation (see e.g. \cite{Levitan,Faddeev,Book-ZS}).

 It is worth mentioning that the double Darboux transformation
with eliminating or adding one bound state at an arbitrary energy
allows one to avoid the problems with singularities of the
transformation kernel $K(x)$ and, as consequence, to avoid the
problems with singularities of the constructed potentials and
solutions. We assume, that $m(x)$ and $q(x)$ do not lead to the
additional singularities on the tested interval. Let us compare the
formulae for the first-order transformation (\ref{L1}), (\ref{trV})
and for the second-order one (\ref{K-int}), (\ref{V_phase}). One can
see, if the first step procedure is based on any arbitrary solution
${\cal U}$ of the generalized equation (\ref{H}), rather than the
ground state wave function, the superpotential $K=-{\cal U}'/{\cal
U}$ becomes singular. Singularities are localized at the zeros of
exited wave functions. It leads to singularities of constructed
potentials and solutions. Unlike  the 1st order transformations the
transformation kernel $K(x)$ of the 2nd order (\ref{K-int}) has no
singularities at zeros of exited wave functions. It means that one
can make transformations on an arbitrary bound state ( not only on
the ground state) and construct the potentials and corresponding
solutions without additional singularities.

It is interesting to note that by using the double
 Darboux transformations (\ref{V_phase}) - (\ref{eta2}) we can construct new potentials $v_2$
without changing the spectrum of the initial potential $v$, i.e.
{\sl fully isospectral potentials}. Indeed, if the bound state
$\lambda_1$ belongs to the spectrum of the initial Hamiltonian
${\cal H}$ and $\Gamma=N_2^2-N^2$ is a difference between the
normalization constants of the bound state $\lambda_1$
 for ${\cal H}_2$ and ${\cal H}$, the formulae (\ref{V_phase}) - (\ref{eta2}) give us  a family of
isospectral potentials and corresponding solutions, since the
normalization constants can be chosen arbitrary. In quantum
mechanics potentials whose spectra coincide and differ only in the
normalizations factors $N_2$ and $N$ of bound states are called
phase-equivalent potentials. Note, the phase-equivalent potentials
have a different shape. They can be more deeper and narrow or more
shallow and wider and possess the same spectral data, except for
normalization constants.

  If we assume, the
transformation function ${\cal U}_1(x)$ to be taken at the energy of
the bound state, which we would like to add to the initial spectrum,
and $\Gamma=N_{2}^2$ is the corresponding normalization constant,
then the formulae (\ref{V_phase})  give us the possibility to
construct a family of two-parametric potentials with a new bound
state $\lambda_1$ and an arbitrary $\Gamma$, whereas the other
spectral characteristics of the spectra produced by the potentials
$v_{2}(x)$ and $v(x)$, coincide.
 Note, constructed potentials $v_{2}(x)$ from this family are isospectral among themselves,
 since posses the coinciding spectra
 and differ only by the normalization constants.

\section{Application}

{\bf The 1-st example.} As an illustrative example we present the
transformed potential and solutions corresponding to the first,
second and third-order Darboux transformations. We start with the
generalized Schr\"odinger equation (\ref{1}) with the repulsive
coulomb potential $v(x)=1/(4~x)$
 \begin{equation} \label{A-1}
  -  \left[
  \frac{d}{dx}\left( \frac{1}{m(x)}\right)\frac{d}{dx}\right]\phi(x)+\frac{1}{4x}
\phi(x)=q(x){\cal E}\phi(x)~,\end{equation}  where we choose the
effective mass as $m=1/x$ and $q=x$.
 The
general solution of this equation can be written as
\begin{equation} \label{A-2}\phi(x)=\frac{C_1~ \sin(kx)}{k
\sqrt{x}}+\frac{C_2~\cos(kx)}{k \sqrt{x}}~. \end{equation}
 Now we
would like to generate potentials with one bound state at the energy
${\cal E}_1=-\kappa_{1}^{2}$ and obtain corresponding solutions by a
first-order supersymmetry transformation (\ref{tr-wV}) and
(\ref{tr-wphi}) applied to a special case $\eta=\eta_1$ of the
general solution (\ref{A-2})
$$\eta_1=\frac{C~\cosh( \kappa_1
x)}{\kappa_1 \sqrt{x}}.$$
 We obtain the transformation operator
$\widetilde{K}_1=\eta_{1}'/\eta_1$ in the form
$$\widetilde{K}_1=-\frac{1}{2x}+\kappa_1\tanh{\kappa_1~x},$$
the potential $v_{1}$ and corresponding solutions $\phi_{1}$ at
${\cal E}\ne{\cal E}_1$
\begin{equation} \label{A-v1}
v_1(x)=\frac{1}{4x}-2 x \kappa_{1}^{2}\left(1-\tanh^2(\kappa_1
x)\right) =\frac{1}{4x}-\frac{2 x
\kappa_{1}^{2}}{\cosh^2(\kappa_1~x)},
 \end{equation}
\begin{equation} \label{A-phi-1}
\phi_1=\left[-\frac{d}{dx} + \widetilde{K}_1 \right]{\phi}
=\left[-\frac{d}{dx}-\frac{1}{2x}+\kappa_1\tanh{\kappa_1~x}\right]{\phi}~.
 \end{equation}
The solution at the energy of transformation ${\cal
E}=\lambda_1=-\kappa_{1}^{2}$ is defined in accordance with
(\ref{tr-U}) as
\begin{equation} \label{A-bound}
{\cal U}=\sqrt{\frac{m}{q}}~\frac{1}{\eta_1
}~=\frac{\sqrt{\kappa_1}}{\sqrt{x}\cosh(\kappa_1~x)}
\end{equation}
and corresponds to the bound state. Note, by varying $\phi$ in
(\ref{A-phi-1}) we recover all solutions of (\ref{1})  with the
transformed potential (\ref{A-v1}) and with $m=1/x$ and $q=x$. In
the case when $\phi$ is chosen to be $\phi(x)=\frac{C\sin(kx)}{k
\sqrt{x}}$ we obtain
\begin{equation} \label{A-phi-1a}
\phi_1=-\frac{ C~\cos(k x)}{\sqrt{x}}+\frac{C~
\kappa_1\tanh(\kappa_1 x) \sin(kx)}{k \sqrt{x}}.
\end{equation}
If  $\phi$ is chosen as $\phi=\frac{C~\exp(\pm ikx)}{k \sqrt{x}}$ we
get the following partial solutions
\begin{equation} \label{A-phi-1b}
\phi_{1}^{\pm}=\left(\mp ik+\kappa_1\tanh(\kappa_1
x)\right)\frac{C~\exp(\pm ikx)}{k \sqrt{x}}~.
\end{equation} Thus, we
presented the simplest example of exactly solvable problem for the
generalized equation (\ref{1}) with $q(x)=x$, $m(x)=1/x$ and with a
real potential (\ref{A-v1}), which is singular at zero. The
potentials, obtained  at different energies of transformation, are
depicted in Fig.1a.

Now employing Darboux transformations of the second-order,
  we shall construct potentials
and solutions of the generalized equation with two bound states. We
define the auxiliary functions $\eta_1$ and $\eta_2$ as follows
$$\eta_1=\frac{\cosh( \kappa_1
x)}{ \sqrt{\kappa_1~x}},~~~~~ \eta_2=\frac{\sinh( \kappa_2 x)}{
\sqrt{\kappa_2~x}}.$$ For the second step we have
$\widetilde{K}=\widetilde{K}_1+\widetilde{K}_2$, where
\begin{eqnarray}
\widetilde{K}_1=\frac{\eta_1^{'}}{\eta_1},~~\widetilde{K}_2=\frac{\chi'_1}{\chi_1},~~
\chi_1=\frac{1}{\sqrt{q~m}}\left(-\eta'_2+\widetilde{K}_1\eta_2\right),~~~\widetilde{K}=-\frac{d}{dx}\left(\ln
\frac{W_{12}}{\sqrt{q~m}}\right). \nonumber
\end{eqnarray}
 The transformed potential $v_2$ having
two bound states at energies $ \lambda_1=-\kappa_{1}^{2}$ and
$\lambda_2=-\kappa_{2}^{2}$ can be written as
\begin{equation} \label{A-v2}
v_2=v-\frac{2\sqrt{q}}{\sqrt{m}}\frac{d}{dx}\left(\frac{\widetilde{K}}{\sqrt{q~m}}\right)-2
~\sqrt{\frac{q}{m}}
\frac{d}{dx}\left[\frac{1}{q}\frac{d}{dx}\left(\sqrt{\frac{q}{m}}~\right)\right].
\end{equation}
Finally, for our choice of $v, ~m$ and $q$ we obtain
\begin{equation}
\label{A-v2-fin} v_2=\frac{9}{4x} -2 x \frac{d^2}{dx^2} \ln
W_{1,2}~,
\end{equation}
where $ W_{1,2}=W(\eta_1,\eta_2)=\frac{1}{x\sqrt{\kappa_2 \kappa_1}
} \Bigl( \kappa_2 \cosh(\kappa_1 x)\cosh(\kappa_2 x)-\kappa_1
\sinh(\kappa_2 x) \sinh(\kappa_1 x)\Bigr) $ and the corresponding
solutions are
$$
\phi_2=\frac{ \sqrt{\kappa_1~x}}{\cosh( \kappa_1
x)}\left(\frac{d}{dx}W_{1,\cal{E}}- \frac{d\Bigl(\ln
W_{1,2}\Bigr)}{dx}W_{1,\cal{E}}\right)~,
$$
where $ W_{1, \cal{E}}=\frac{C}{x~k \sqrt{\kappa_1} }
\Bigl(k~\cosh(\kappa_1 x)\cos(k x)-\kappa_1\sinh(\kappa_1 x)
\sin(kx) \Bigr)$ if $\phi(x)$ is chosen as $\phi(x)=\frac{C~
\sin(kx)}{k \sqrt{x}}$.  We put $\kappa_1<\kappa_2$. The potential
having two bound states are depicted in Fig.1c.

By using (\ref{VM-pr}) one can construct the potential $v_3$ for the
generalized Schr\"odinger equation (\ref{H-1}) having three bound
states
\begin{equation}\label{A-VM}
 v_3=\frac{13}{4 x}-2 x\frac{d^2}{dx^2}\ln W_{1,2,3} ~,
 \end{equation}
where the auxiliary functions $\eta_1$, $\eta_2$ and $\eta_3$
determined as
$$\eta_1=\frac{\cosh( \kappa_1
x)}{\sqrt{\kappa_1 ~x}},~~~~~ \eta_2=\frac{\sinh( \kappa_2 x)}{
\sqrt{\kappa_2~x}},~~~\eta_3=\frac{\cosh( \kappa_3 x)}{
\sqrt{\kappa_3~x}}$$ give us the Wronskian $W_{1,2,3}$
\begin{eqnarray}
 W_{1,2,3}&=& \frac{1}{x^{3/2}\sqrt{\kappa_1\kappa_2\kappa_3}}\Bigl[
\cosh(\kappa_{1} x ) \cosh (\kappa_{2} x ) \kappa_{2} \sinh(\kappa_3
x ) \kappa_{3}^{2}-\cosh(\kappa_{1} x)
\sinh(\kappa_2 x ) \kappa_{2}^{2} \cosh ( \kappa_3 x ) \kappa_3 -\nonumber\\
&- &\sinh (\kappa_1 x ) \kappa_{1}\,\sinh ( \kappa_2 x ) \sinh(
\kappa_3 x ) \kappa_{3}^{2}+ \sinh (\kappa_1 x ) \kappa_1\,\sinh
(\kappa_3 x)
 \sinh (\kappa_2 x ) \kappa_{2}^{2}+\nonumber\\
& + &\cosh (\kappa_1 x) \sinh (\kappa_2 x ) \kappa_{1}^{2}\cosh
(\kappa_3 x ) \kappa_3 -\cosh( \kappa_1 x )
\kappa_{1}^{2}\cosh(\kappa_2 x )\kappa_2 \sinh(\kappa_3 x)
 \Bigr]~.
\end{eqnarray}
\begin{figure}
\vspace{45mm}
\makebox[46mm][l]{\includegraphics{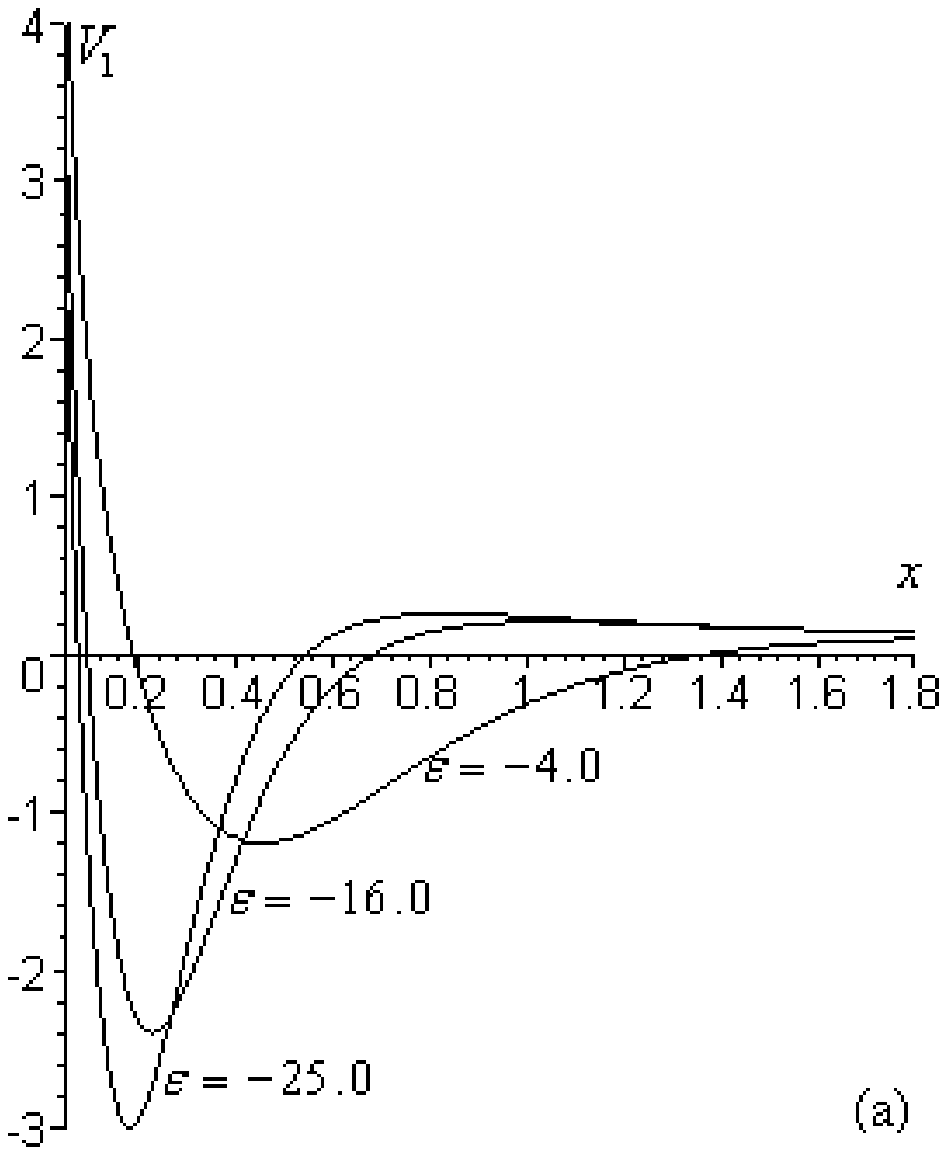}}%
\makebox[5mm]{}%
 \makebox[46mm][l]{\includegraphics{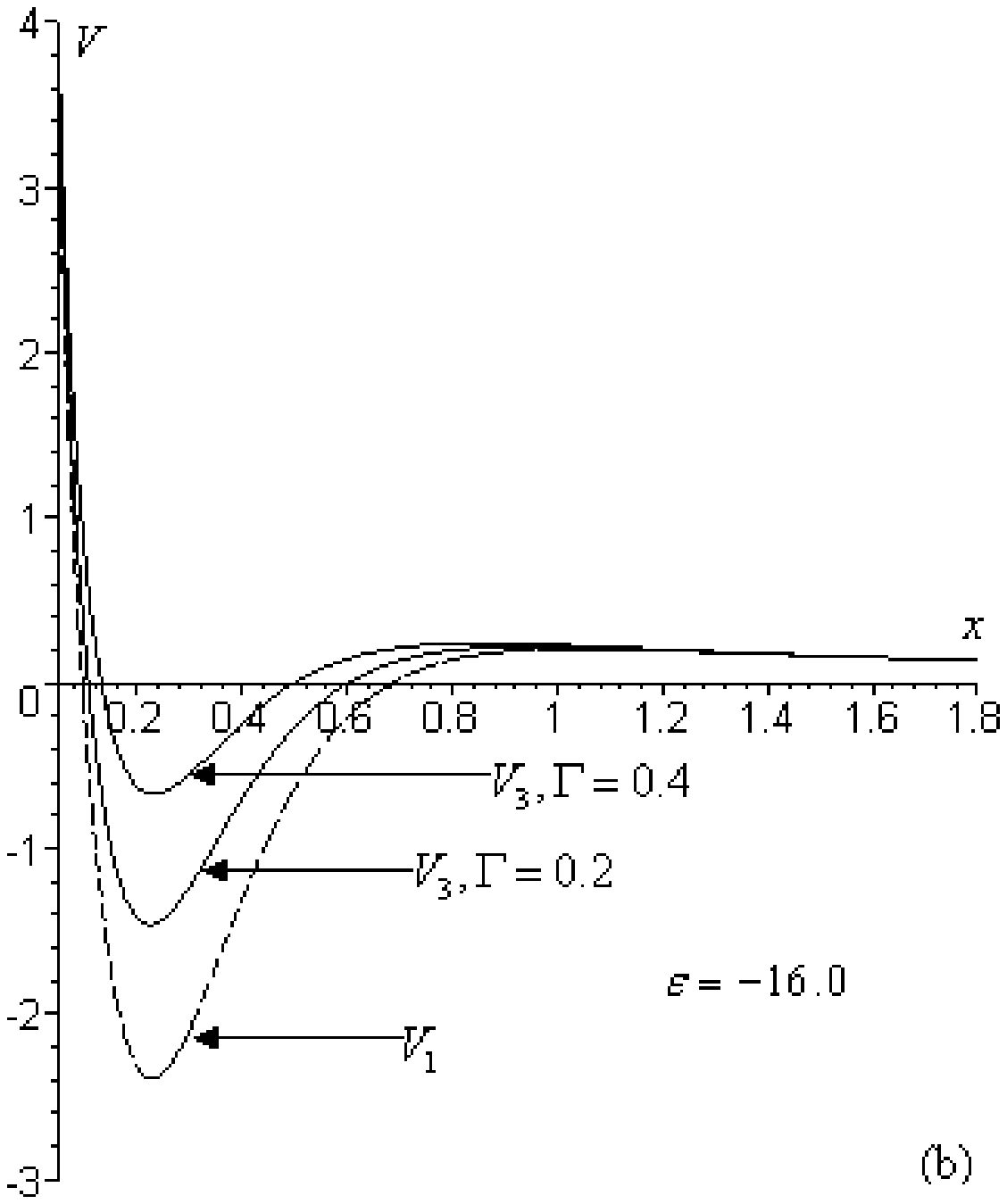}}%
\makebox[5mm]{}%
\makebox[46mm][l]{\includegraphics{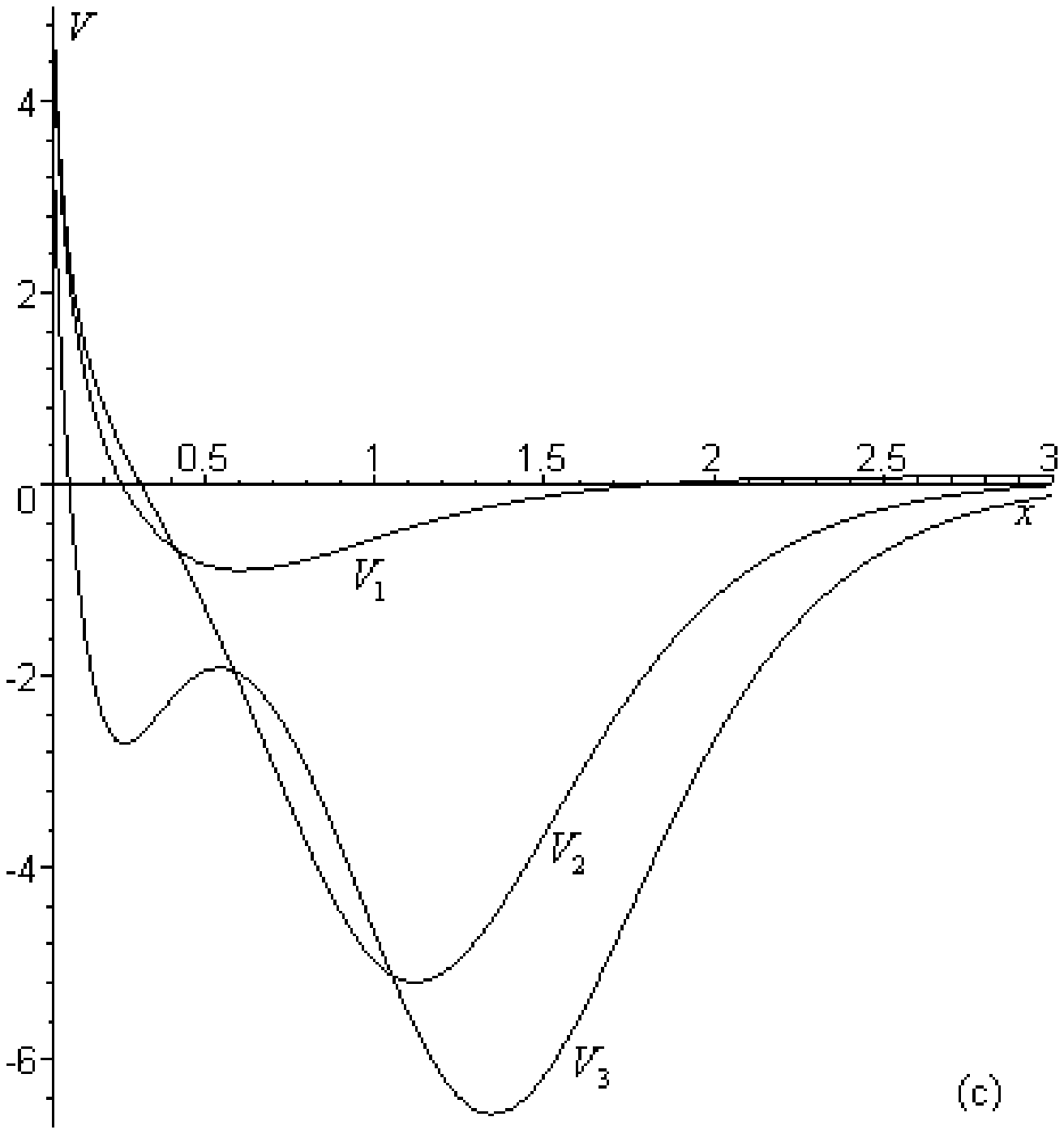}}%
\caption{(a) Potentials $v_1(x)$ corresponding to one bound state
created at different energies. ~(b) Completely isospectral
potentials with bound state energy ${\cal E}_1=-16.0$, dashed line
corresponds to the initial potential $v_1$, solid lines correspond
to its isospectral  potentials.~(c) Potentials $ v_{n},~n=1,2,3 $
having one, two and three bound states, respectively. }
\end{figure}
 As an illustrative example we present the potentials $v_1$,
obtained at the energy of transformation ${\cal E}_1=-4$, $v_2$
obtained at the energies  ${\cal E}_1=-4,~{\cal E}_2=-16$ and $v_3$
calculated at the energies of transformation ${\cal E}_1=-4,~{\cal
E}_2=-16,~{\cal E}_3=-25$. They are depicted in Fig.1c.

Now  we are going to  construct {\sl isospectral potentials}. As an
initial potential we take the potential $v_1$, obtained within the
first-order intertwining (\ref{A-v1}). Using (\ref{V_phase}) and
changing  $v$ by $v_1$ from (\ref{A-v1}) and ${\cal U}_1$ by ${\cal
U}$ from (\ref{A-bound}) after simplification we obtain a family of
potentials with the same eigen-value and different values $\Gamma$
\begin{equation}
\label{Ph-3}
v_3(x)=\frac{1}{4x}-\frac{2x\kappa_{1}}{\cosh^2(\kappa_1~x)}-2x\frac{d}{dx}\ln\left[1+\Gamma\tanh\kappa_1~x\right]
 \end{equation}
The initial potential $v_1$ and its isospectral potentials
(\ref{Ph-3}) are presented in Fig.1b.

{\bf  The 2-nd example.} As the following  example let us consider
creation of new bound states {\sl for effective mass Schr\"odinger
equation}
\begin{equation}
 \label{B-0} -\left[\frac{d}{dx}\left(\frac{1}{m(x)}\right)\frac{d}{dx}\right]
\phi(x)+v(x)\phi(x)={\cal E}~\phi(x)
 \end{equation}
 that corresponds to the generalized equation (\ref{1}) with
$q(x)=1$. We  choose the effective mass in the form
$m(x)=\alpha^2/x^2$, the initial potential $v(x)=0$ and start with
the equation
\begin{equation}
 \label{B-1} -\left[\frac{d}{dx}\left(\frac{1}{m(x)}\right)\frac{d}{dx}\right]
\phi(x)={\cal E}~\phi(x).
 \end{equation}
The general solution of equation(\ref{B-1}) can be written as
\begin{equation} \label{B-2}\phi(x)=\frac{\alpha}{\sqrt{x}}\left[c_1~ \sin(\alpha\nu\ln(x))
+c_2\cos(\alpha\nu\ln(x))\right], \end{equation} where $c_1,~c_2$
are free constants and $\nu^2=(-1+4\alpha^2k^2)/4\alpha^2$. Recently
in \cite{suz-shul} we have constructed the potentials for effective
mass Schrodinger equation (\ref{B-0}) with creation of one and two
bound states without investigation of potential forms. Here we would
like to apply our technique to construction of double-well and even
triple-well potentials  with creation of two and three bound states,
to generation of completely isospectral potentials and to
investigation of the influence of position dependent mass on the
form of constructed potentials.

{\sl Construction of completely isospectral potentials.}
 As an
initial potential we take the potential $v_1$, obtained in
\cite{suz-shul} within the $1$-st order intertwining
\begin{equation}\label{V-2}
 v_1=-\frac{2\gamma^2}{\cosh^{2}\left(\alpha\gamma\ln(x)\right)}
 . \end{equation}
It can be easily obtained from (\ref{tr-wV}) at q(x)=1 with a
particular solution of  $\eta_1$  given as $ \eta_1(x) =
\sqrt{\frac{\alpha}{x}}~\cosh\left(\alpha\gamma_1\ln(x) \right)~.$
 The solution at energy of transformation ${\cal E}_1=-\kappa_{1}^{2}$ is
\begin{equation}\label{U-2}
{\cal U}=\frac{\sqrt{
m^*(x)}}{\eta_1(x)}=\frac{\sqrt{\alpha}}{\sqrt{x}
\cosh\left(\alpha\gamma\ln(x)\right)}.
\end{equation}
and corresponds to the bound state.
 Using (\ref{V_phase}) and
replacing   $v$ by $v_1$  from  (\ref{V-2}) and ${\cal U}_1$ by
${\cal U}$ from (\ref{U-2}) after simplification we obtain
two-parametric family of isospectral potentials
\begin{eqnarray}\label{V3-isosp}
v_3 =-\frac{2\gamma^2}{\cosh^{2}\left(\alpha\gamma\ln(x)\right)}
-\frac{2x}{\alpha}\frac{d}{dx}\left[\frac{x}{\alpha}\frac{d}{dx}\ln
P\right]~,
\end{eqnarray}
where
$$P=1+\Gamma\int^x\frac{\alpha^2}{x^2\cosh^{2}\left(\alpha\gamma\ln(x)\right)}dx~.$$
All these potentials $v_3$ posses a single bound state each with the
same energy ${\cal E}_1=-\kappa_{1}^{2}$ as the initial potential
$v_1$, as well as the normalization constants including in $\Gamma$
can be chosen arbitrary.  In Fig.2a we have plotted the initial
potential calculated by the formula (\ref{V-2}) and its strictly
isospectral potentials, calculated by (\ref{V3-isosp})  at different
$\Gamma$ .
\\
{\sl Influence of distance between levels and effective mass on the
form of potentials.}
 By using second-order intertwining let us construct a potential with creation of two bound
states. For this we define the auxiliary transformation functions as
follows
\begin{eqnarray}
\label{eta_12} \eta_1(x) =
\sqrt{\frac{\alpha}{x}}~\cosh\left(\alpha\gamma_1\ln(x) \right), ~~~
\eta_2(x) =
\sqrt{\frac{\alpha}{x}}\sinh\left(\alpha\gamma_2\ln(x)\right),
\end{eqnarray}
where $\gamma_{i}^2=-(1+4\alpha^2\kappa_{i}^2)/4\alpha^2,~~i=1,2 $.
We put $\kappa_1<\kappa_2$. The potential $v_2$, obtained within the
second-order Darboux transformation, can be  expressed from
(\ref{A-v2}) with $q(x)=1$
\begin{eqnarray}\label{V2x}
v_2=v-\frac{2}{\sqrt{m}}\frac{d}{dx}\left[
\frac{1}{W_{12}}~\frac{d}{dx}\frac{W_{12}}{\sqrt{m}}\right]-\frac{2}{\sqrt
{m}}\frac{d^2}{dx^2}\frac{1}{\sqrt {m}}~.
\end{eqnarray}
This formula coincides with expression obtained in \cite{suz-shul}
for effective mass Schr\"odinger equation. For our choice of $m(x)$
the last term vanishes and the potential is written as
\begin{eqnarray}\label{V2-mx}
v_2=-\frac{2x}{\alpha}\frac{d}{dx}\left[ \frac{x}{\alpha}
~\frac{d}{dx}\ln{W_{12}}\right]~,
\end{eqnarray}
where Wronskian is determined as
$$W_{12}=\frac{\alpha^2}{x^2}\Bigl(\gamma_2\cosh(\alpha\gamma_2\ln
(x)) \cosh(\alpha\gamma_1\ln(x))-\gamma_1\sinh(\alpha\gamma_1\ln
x)\sinh(\alpha\gamma_2\ln(x))\Bigr).$$ The potentials having two
bound states are presented in Fig.2~b,c.
\begin{figure}
\vspace{45mm}
\makebox[46mm][l]{\includegraphics{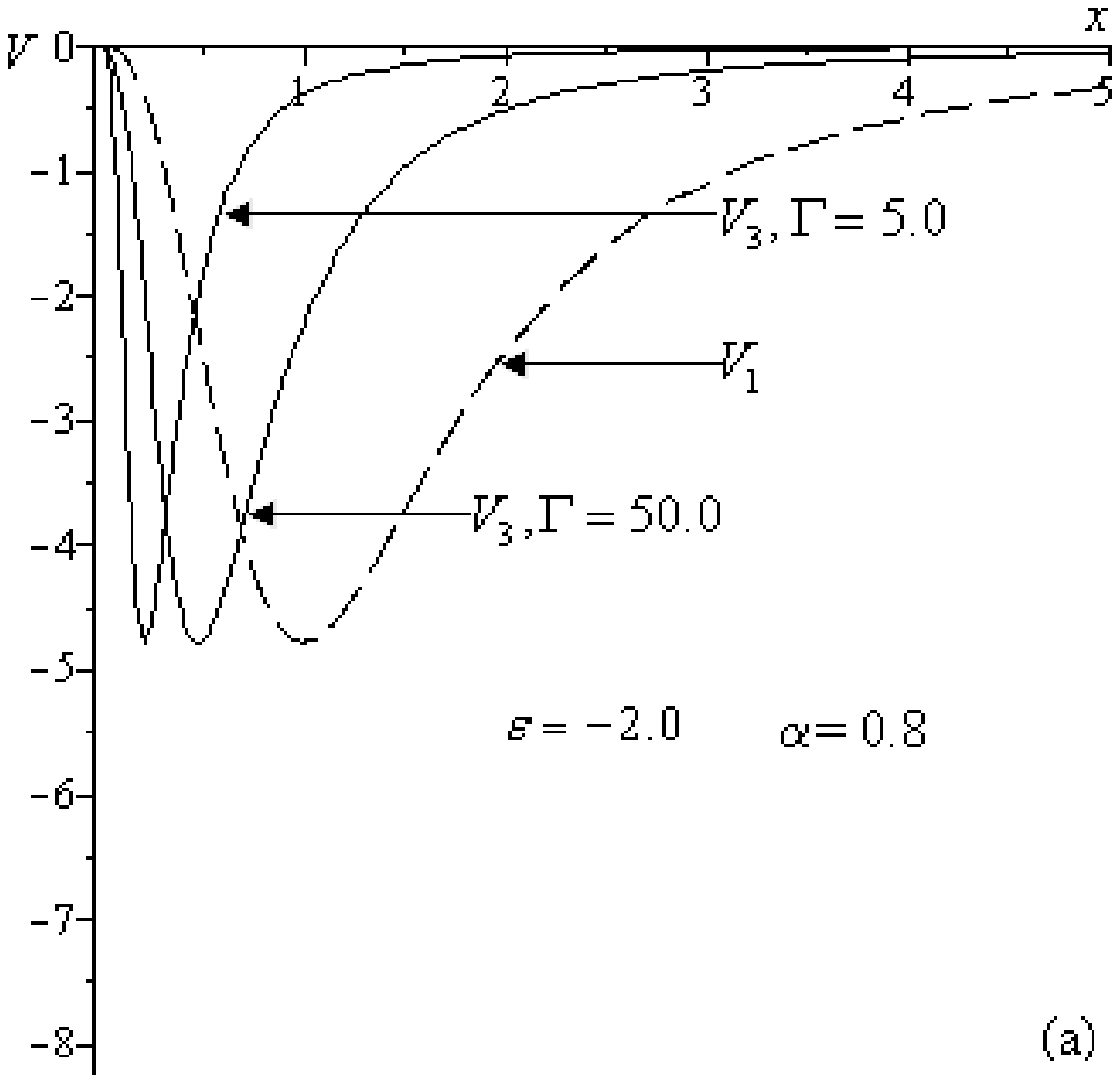}}%
\makebox[5mm]{}%
 \makebox[46mm][l]{\includegraphics{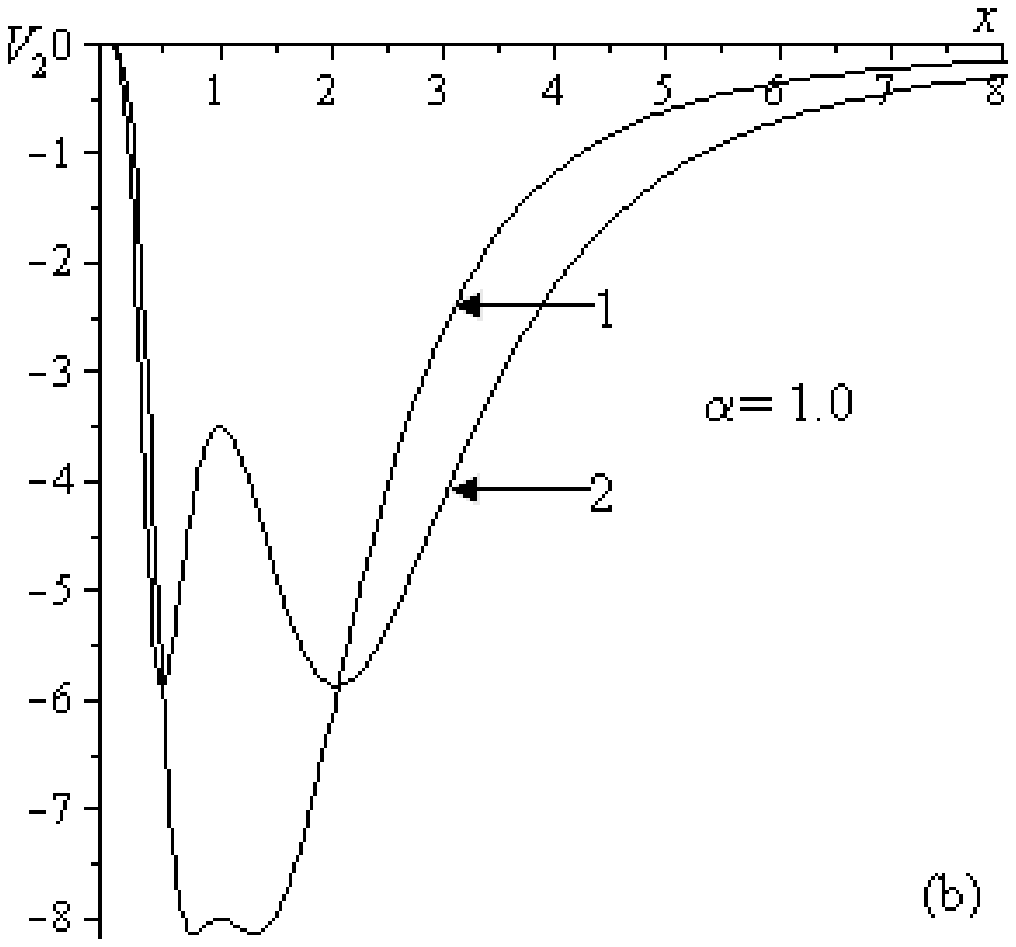}}%
\makebox[5mm]{}%
\makebox[46mm][l]{\includegraphics{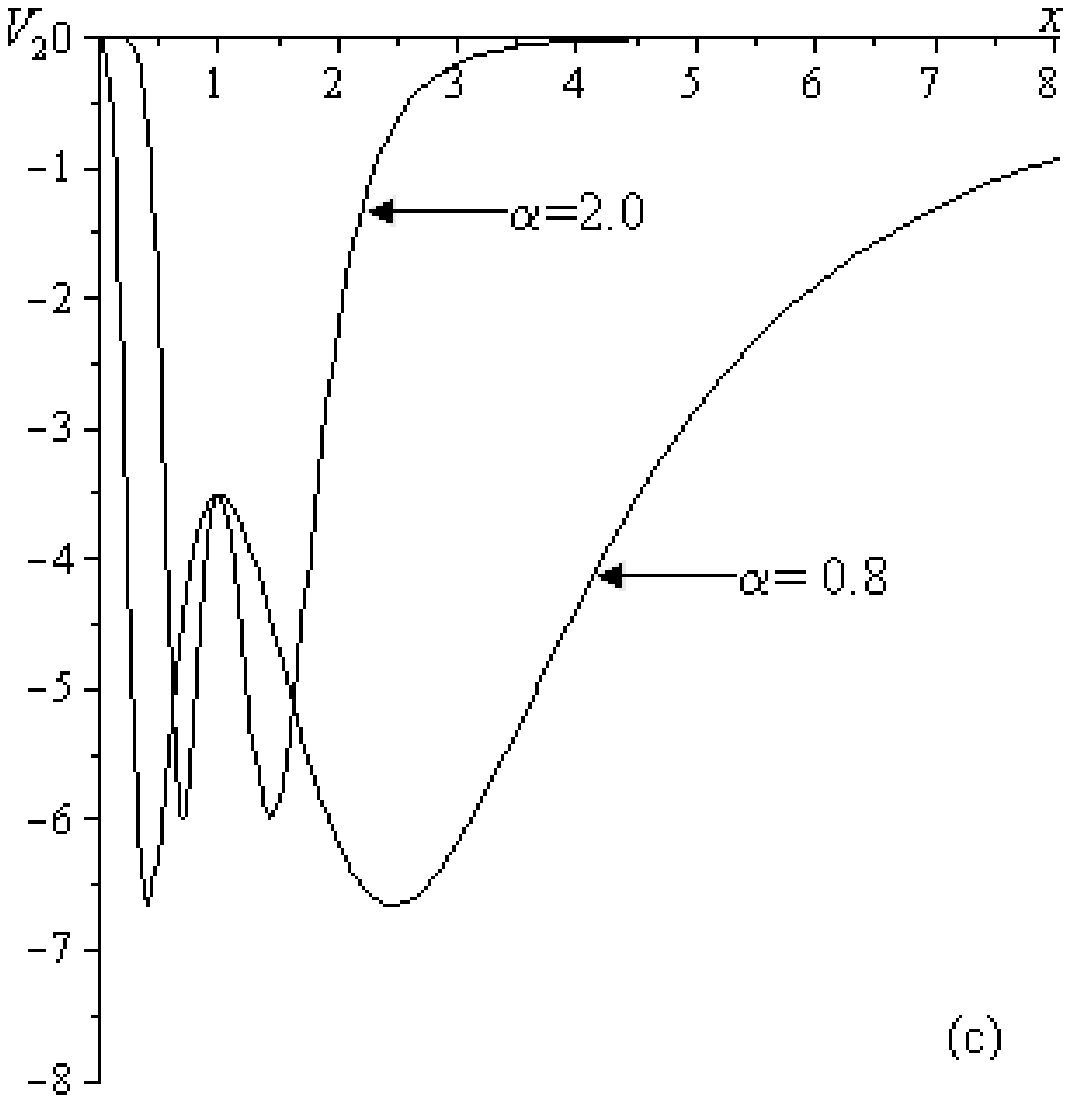}}%
\caption{ (a) Completely isospectral potentials with bound state
energy ${\cal E}_1=-2.0$, dashed line corresponds to the initial
potential $v_1$, solid lines correspond to isospectral partner
potentials.~(b) The change in form of potentials $v_2(x)$ with two
bound states as the levels approach each other: \break $~1. \to
{\cal E}_1=-2.0, ~{\cal E}_2=-6.0, ~ 2. \to {\cal E}_1 =-2.0,~{\cal
E}_2=-3.75$.~(c)~The influence of $m(x)$ on the behavior of
transformed potentials $v_2(x)$ with two bound states ${\cal E}_1
=-2.0,~{\cal E}_2=-3.75$.}
\end{figure}
The graphs in (Fig.2~b) depict the forms of constructed potentials
in dependence from the distance between energy levels. One can see
if the levels are close to each other, we construct asymmetric
double well potentials (curves  in Fig.2~2b) another, we construct
asymmetric potentials (Fig.2~1b). Note,  double well potentials have
attracted some attention over the last years (see, e.g.,
\cite{Sukhatme,Novaes,Tralle}).  Asymmetric double well potentials
for the ordinary Schr\"odinger equation were investigated in
\cite{Sukhatme} with introducing a special parameter of asymmetry.
In our case asymmetry in forms is a consequence  of the
position-dependent mass $m(x)$ that is singular at zero. The
influence of $m(x)$ on the form of constructed potential $v(x)$ is
demonstrated in Fig.2~c,b.
\begin{figure}
\vspace{40mm}
\makebox[46mm][l]{\includegraphics{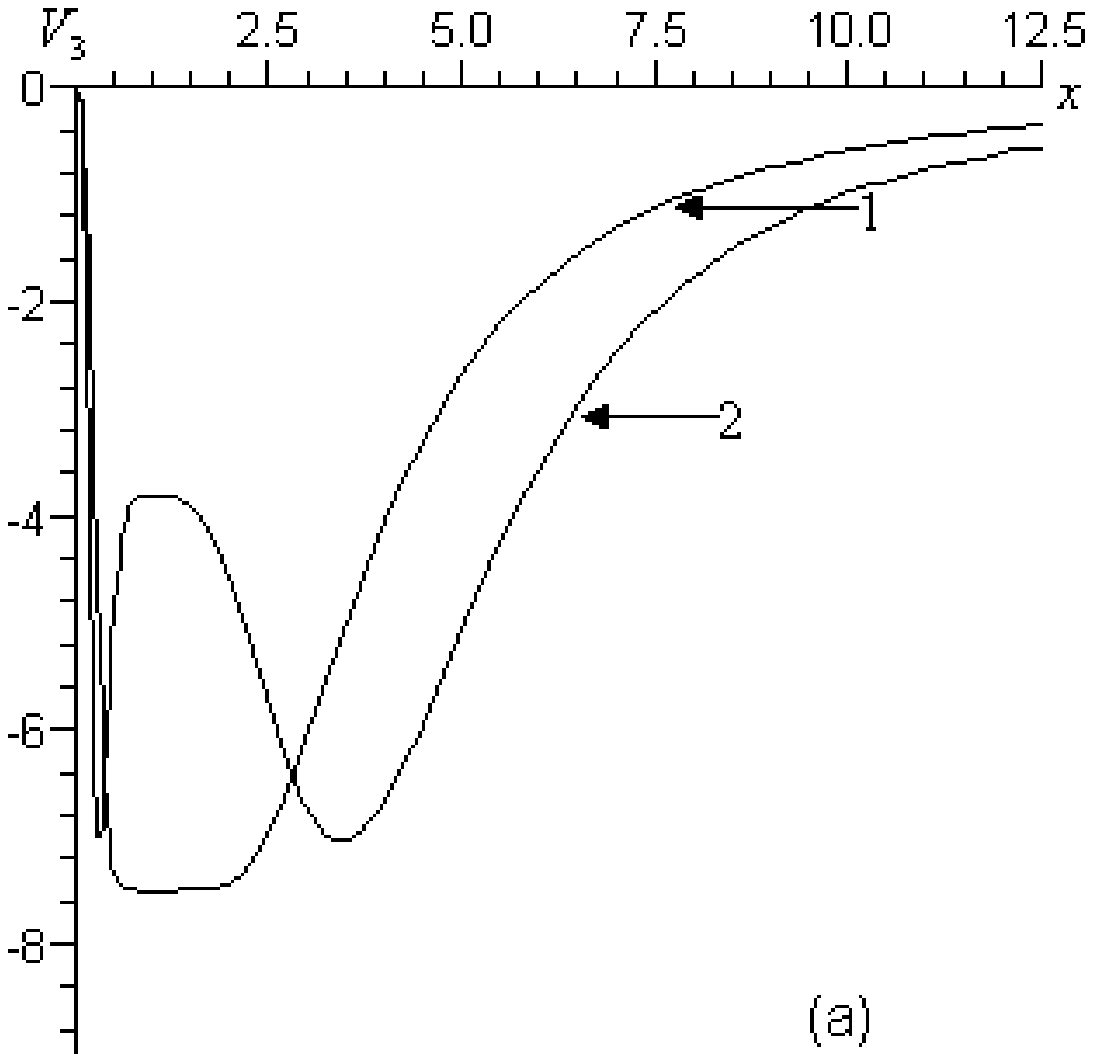}}%
\makebox[5mm]{}%
 \makebox[46mm][l]{\includegraphics{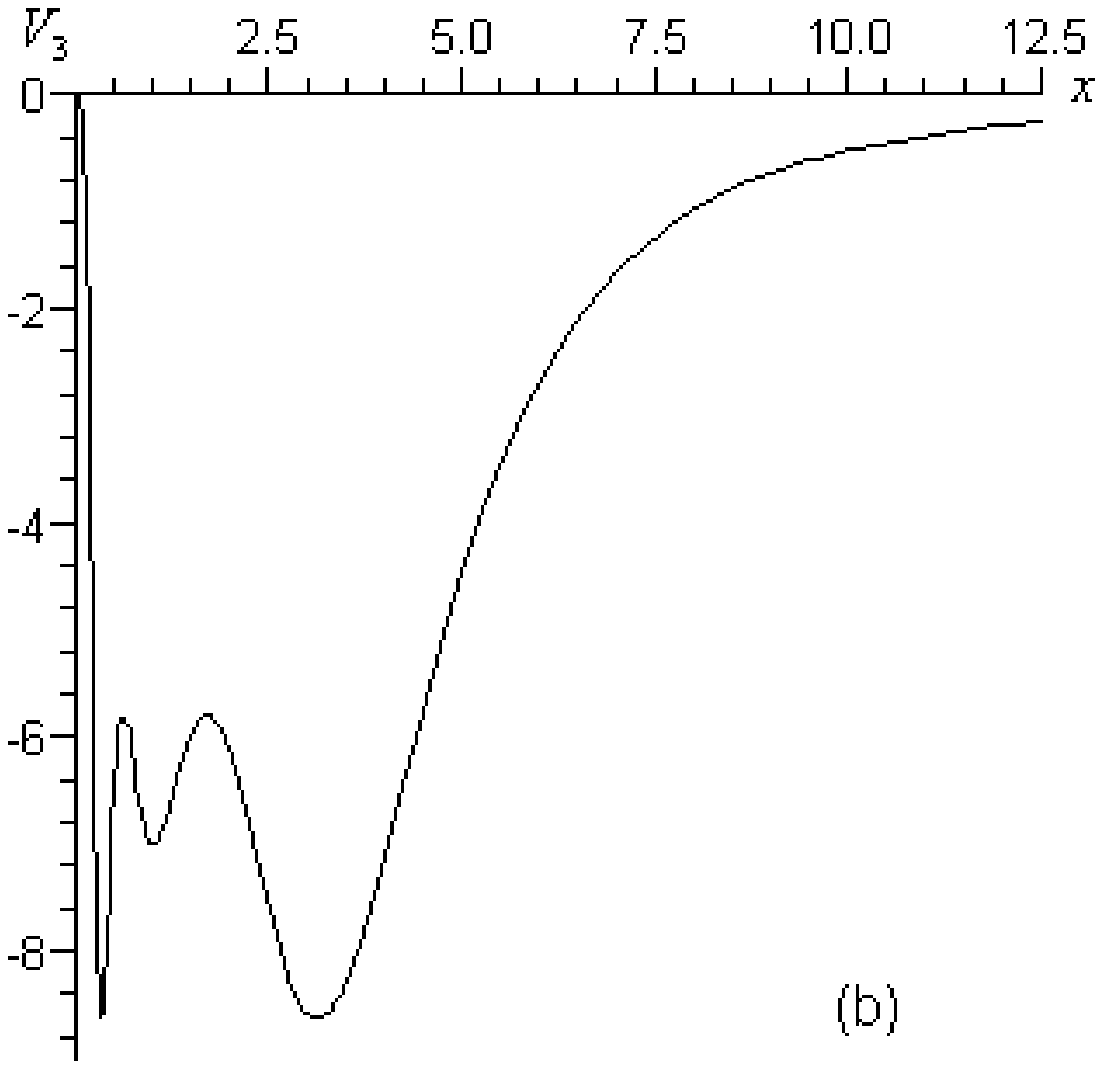}}%
\makebox[5mm]{}%
\makebox[46mm][l]{\includegraphics{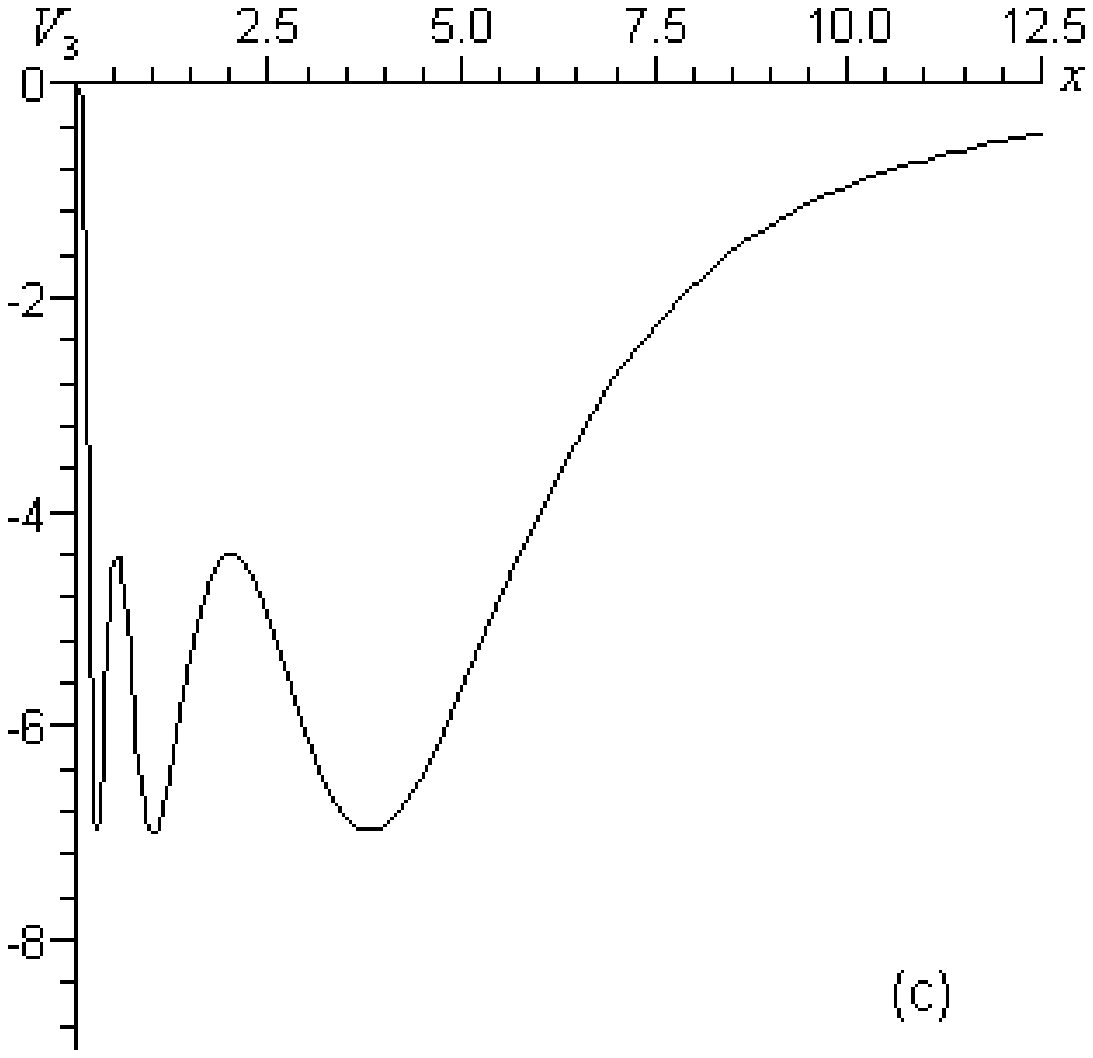}}%
\caption{ The change in form of potentials $V_3(x)$ with tree bound
states as the levels come close to each other: \break $~(a)~1. \to
{\cal E}_1=-1.0,~{\cal E}_2=-3.75,~ {\cal E}_3=-6.25,~~ 2. \to {\cal
E}_1 =-1.0,~{\cal E}_2=-3.75,~ {\cal E}_3=-5.0;$ ~$(b)~ {\cal E}_1
=-2.0,\break~{\cal E}_2=-4.75,~ {\cal E}_3=-6.0;$~ $(c)~~{\cal E}_1
=-2.0,~{\cal E}_2=-3.75,~ {\cal E}_3=-5.0$.}
\end{figure}
Our analysis shows that the larger $\alpha$, the shallow and narrow
constructed potential (see Fig. 2~b, curve 2 and Fig. 2~c). Another
words, increasing  $m(x)$ leads to decreasing  potential $v(x)$.

The next considered example illustrates the possibility to construct
potentials having three bound states, for all that  we can generate
double well potentials and even triple well potentials. Employing
the third-order Darboux transformations (\ref{Sv-n}) with $q(x)=1$,
the potential $v_3$ can be written as
\begin{eqnarray}\label{V3x}
v_3 =-\frac{2x}{\alpha}\frac{d}{dx}\left[ \frac{x}{\alpha}
~\frac{d}{dx}\ln{W_{123}}\right]~.
\end{eqnarray}
 The Wronskian $ W_{1,2,3}$ is
determined by the auxiliary functions $\eta_1(x),~ \eta_2(x),
~\eta_3(x)$, where $\eta_1(x)$,~$\eta_2(x)$, is defined in
(\ref{eta_12}) and $\eta_3$ is given as $\eta_3(x) =
\sqrt{\frac{\alpha}{x}}~\cosh\left(\alpha\gamma_3\ln(x) \right).$
The potentials $v_3$ calculated by the formula (\ref{V3x}) are
plotted in Fig.3. As in the previous case with two bound states, the
forms of constructed potentials depend on the space between energy
levels. We can see if levels  are sufficiently distant from one
another, we construct simple asymmetric potentials presented in
Fig.3~a (curve 1), if two levels out of three are close to each
other, we construct asymmetric double well potentials (Fig.3~a,
curve 2), if three levels are close to each other, we construct
asymmetric triple well potentials (Fig.3~b, and Fig.3~c). As a final
remark, let us note that different distances between levels give us
different shapes of potentials. It can be very important for
construction and investigation of quantum systems with needed
spectral properties, e.g. in nanoelectronics \cite{Tralle}.

\section*{Conclusion}
By application of the intertwining operator technique to generalized
Schr\"odinger equation with position dependent mass and with energy
dependent potentials,  Darboux transformations of an arbitrary order
have been constructed. It has been shown that $n$th order Darboux
transformation is equivalent to the resulting action of a chain of
first-order Darboux transformations. Our generalized Darboux
transformations comprises the position-dependent effective mass case
and the case of linearly energy-dependent potentials, as well as the
conventional case of Schr\"odinger equation. The integral Darboux
transformation method has been elaborated for the generalized
Schr\"odinger equation.
 An interrelation has been found between the differential
and integral transformations.
  The integral Darboux transformations have been applied to
 generation of isospectral   Hamiltonians differing by one
and by two bound states from the spectrum of the initial one.  It
has been shown how to produce completely isospectral   Hamiltonians
to a given initial one. On concrete examples it has been
demonstrated how to apply the Darboux transformation technique for
modeling quantum well potentials  with the given spectrum.
Hamiltonians with different number of levels have been produced and
the influence of the distance between levels on the shape of
constructed potentials has been investigated, in particular,
asymmetric double well and triple well potentials have been  built.
The influence of the position-dependent mass on the behaviour of
constructed potentials has been studied, too.

\section*{Acknowledgments} This work was supported in part  by a grant of the Russian
Foundation for Basic Research 09-01-00770.


\begin{thebibliography}{99}
\bibitem{Schr}E. Schr\"odinger,  Proc.Roy.Irish. Acad., A. {\bf 46} (1940)
p.9; A. {\bf 47}, 53 (1941).
\bibitem{Darboux}
 M.G. Darboux, Comptes Rendus Acad. Sci. Paris. {\bf 94}, 1343
(1882);  {\bf 94}, 1456 (1882).
\bibitem{Witten}E. Witten, Nucl.Phys. B {\bf 185} (1981) 513; B {\bf 202}, 253 (1982).
\bibitem{Bagrov}  V.G. Bagrov, D.M. Gitman, Exact Solutions of
Relativistic Wave Equations, (Kluwer Academic Publishers, Dordrecht/
Boston/ London) 1990, 323p.
\bibitem{Chadan}
{\it  K. Chadan, P. C. Sabatier}, "Inverse Problems in Quantum
Scattering Theory", 2nd edn (New York: Springer) 1989, 499p.
\bibitem{Book-ZS} B.N. Zakhariev and A.A. Suzko,
   "Direct and inverse problems, (Potentials in quantum scattering)",
    (New York: Springer) 1990, 223p.
\bibitem{Junker}
{\it Junker G.} "Supersymmetric Method in Quantum and Statistical
Physics",  (New York: Springer), 1996, 173p.
\bibitem{matveev} V.B. Matveev and M.A. Salle,
$"$Darboux transformations and solitons$"$, (Springer, Berlin, 1991)
\bibitem{darbbook} C. Gu, H. Hu and Z. Zhou, $"$Darboux transformations in integrable systems$"$,
(Mathematical Physics Studies 26, Springer, Dordrecht, The
Netherlands, 2005)
\bibitem{Andrianov} A.A. Andrianov, M.V. Ioffe, V. Spiridonov,
Phys. Lett. A {\bf 174},  273 (1993);  A.A. Andrianov, F. Cannata,
J.Phys. {\bf A 37},  10297 (2004).
\bibitem{Amado}
R.D. Amado, F. Cannata and J.P. Dedonder, Phys. Rev. Lett.
 {\bf 61}, 2901 (1988); Phys. Rev.  A{\bf 38}, 3797 (1988);
Int. J. Mod. Phys. {\bf 5},  3401 (1990).
\bibitem{Scripta84}
 B.V. Rudyak, A.A. Suzko, B.N. Zakhariev, Physica Scripta, {\bf 29}, 515 (1984)
\bibitem{Suz85}
A.A. Suzko, Physica Scripta, {\bf 31}  (1985) 447; Physica Scripta,
{\bf 34}, 5 (1986);
\bibitem{Suz93}
A.A. Suzko, Sov. J. Nuclear Physics {\bf 55}, 1359 (1992);
 Sov.J.Part. and Nucl. {\bf 24}, No.4, 485 (1993);
 /in: /{\it
Quantum Inversion Theory and its Applications}, Lect. Notes in
Phys., vol. 427, Ed. H. Geramb, Springer, Berlin, 1993, pp. 67-106.
\bibitem{Gelfand}
 I. M. Gel'fand, B. M. Levitan,
Izv.Akad.Nauk. SSR ser.Math. {\bf 15}, p. 309-360 (1951).
\bibitem{Marchenko}
 Z. S. Agranovich,  V. A.  Marchenko, Inversion Problem of Scattering Theory (Gordon
and Breach, New York) 1963.
\bibitem{Levitan}
B.M. Levitan, Inverse Sturm-Liouville problems, Nauka, Moscow, 1984.
\bibitem{shabat}
V.E. Zakharov, A.B. Shabat, Funct. Anal. Appl. {\bf 8}, 226 (1974);
ibid.
 {\bf 13}, 166 (1979).
\bibitem{pavlov}
B. Pavlov, The Theory of Extensions and explicitly solvable models,
(In Russian) Uspekhi Mat. Nauk, {\bf 42}, 99 (1987).
\bibitem{Acta}
A.A. Suzko,  G. Giorgadze, Physics of Atomic Nuclei, {\bf 70}, 604
(2007); A.A. Suzko,  I. Tralle, Acta Physica Polonioca B, {\bf 39},
No.3, p. 1001-1023 (2008).
\bibitem{Feshbach} H. Feshbach, Ann.Phys.(N.Y.)
{\bf 5}, 357 (1958).
\bibitem{Lipperhide} P.Fr\"obrich and R. Lipperhide, {\it Theory of
Nuclear Reaction} (Clarendon, Oxford, 1996).
\bibitem{Ring} P.Ring and P.Schuck, {\it The Nuclear Many Body
Problem}, Springer, New York, 1980 p.211
\bibitem{Babikov} V.V. Babikov,  {\it Method of Phase function
  in Quntum Mechanics}, Nauka, Moscow, 1976;\\
 S.I. Vinitsky et.al., Physics of Atomic Nuclei, {\bf 64}, 27
 (2001);
\\
 M.I. Jaghoub, Phys. Rev.A {\bf 74},  032702 (2006).
 \bibitem{Arias}
F. Arias de Saavedra et.al, Phys. Rev. B {\bf 50}, 4248 (1994).
\bibitem{Morrow} R.A.  Morrow and K.R. Brownstein,  Phys. Rev. B {\bf 30},  678
(1984).
\bibitem{Einevoll}
 G.T. Einevoll, P.C. Hemmer and J.Thomesn, Phys. Rev. B {\bf 42}, 3485 (1990).
\bibitem{Plastino} A.R. Plastino et al., Phys. Rev. A {\bf 60}, 4318 (1999).
\bibitem{Milanov} V. Milanovi\'{c}, Z. Iconi\'{c},
J.Phys. A: Math.Gen. {\bf 32},  7001 (1999).
\bibitem{Roy}
B. Roy and P. Roy, J.Phys. A {\bf 35},  3961 (2002).
\bibitem{Koca} R.Ko\c{c} and M.Koca,  J.Phys. A {\bf 36},  8105 (2003).
\bibitem{Quesne} C. Quesne, Annals of Physics
{\bf 321}, Issue 5, 1221 (2006).
\bibitem{ss-JPA} A.A. Suzko and A. Schulze-Halberg,  J.Phys. A; Math.Gen. {\bf 42}, 295203 (2009).
\bibitem{suz-shul} A.A. Suzko and A. Schulze-Halberg, Phys.Lett.
A, {\bf 372}, 5865 (2008).
\bibitem{Roy2}Bikashkali Midya, B. Roy, R.Roychoudhury, J.Math.Phys.
{\bf 51}, 022109 (2010).
\bibitem{gener} A.A. Suzko,  A. Schulze-Halberg,
E.P. Velicheva, Physics of Atomic Nuclei, {\bf 72}, 858 (2009).
 \bibitem{Goser} K. Goser,  P. Gl\"{o}sek\"{o}tter, J.
Dienstuhl, {\it Nanoelectronics and Nanosystems. From Transistors to
Molecular and Quantum Devices}. Springer-Verlag, Berlin, 2004.
\bibitem{physE} Special issue of Physica E: Low-dimensional Systems
and Nanostructures, {\bf  14}, No.1/2, 2002
\bibitem{Bastard} G. Bastard, {\it Wave Mechanics
applied to semiconductor heterostructure}(Les Editions de Physique,
Les Ulis, France, 1988).
\bibitem{Sukhatme} F.Cooper, A.Khare, U.Sukhatme, Physics Reports, {\bf 251},
p. 267-385 (1995);\\
Wai-Yee Keung, Eve Kovacs,
U.P. Sukhatme, Phys. Rev. Lett.  {\bf 60}, 41 (1988);\\
 A. Gangopadhyaya, P.K. Panigrahi, U. P.
Sukhatme, Phys. Rev. A, {\bf 47}, 2720 (1993).
\bibitem{Novaes}
M. Novaes, M.A.M. Aguiar, J.E.M. Hornos, J.Phys. A; Math.Gen. {\bf
36},  5773 (2003).
\bibitem{Tralle} K.
Majchrowski, W.Pa\'sko, I. Tralle, Phys. Lett.A, {\bf 373}, 2959.
(2009)
\bibitem{Faddeev} L.D. Faddeev Usp. Mat. Nauk {\bf 14}, 57 (1959).
\end{thebibliography}
\end{document}